\documentclass[twocolumn,apj]{aastex63}
\usepackage{booktabs}
\usepackage{amsmath,amsthm,amsfonts,amssymb,extarrows}
\usepackage{apjfonts}
\usepackage{float}
\usepackage{xcolor}

\definecolor{mBlue}{RGB}{51, 77, 167}

\hypersetup{linkcolor=red,citecolor=blue,filecolor=red,urlcolor=blue}

\newcommand{\angstrom}{\mbox{\normalfont\AA}}

\received{2020 December 23; revised 2021 March 19; accepted 2021 March 23; published 2021 June 11}
\shorttitle{S~\textsc{{{xv}}} line polarization and cross sections}
\shortauthors{Chintan Shah \textit{et al}}% (2021) ApJ \textbf{914} 34}

\begin{document}
	
	\title{\textsc {High-resolution Laboratory Measurements of K-shell X-ray Line Polarization and \\ Excitation Cross Sections in Heliumlike \ion{S}{15} Ions}}
	
	%========= AUTHORS ===================================================
    \author[0000-0002-6484-3803]{Chintan Shah}
	\affil{NASA Goddard Space Flight Center, 8800 Greenbelt Rd, Greenbelt, MD 20771, USA; \href{mailto:chintan@mpi-hd.mpg.de}{chintan@mpi-hd.mpg.de}}
	\affil{Lawrence Livermore National Laboratory, 7000 East Avenue, Livermore, CA 94550, USA}
	\affil{Max-Planck-Institut f\"ur Kernphysik, Saupfercheckweg 1, D-69117 Heidelberg, Germany}
		
	\author[0000-0003-3057-1536]{Natalie Hell}
	\affil{Lawrence Livermore National Laboratory, 7000 East Avenue, Livermore, CA 94550, USA}
	
	\author[0000-0002-4982-9413]{Antonia Hubbard}
	\affil{Lawrence Livermore National Laboratory, 7000 East Avenue, Livermore, CA 94550, USA}
	
	\author[0000-0001-9136-8449]{Ming Feng Gu}
	\affil{Space Science Laboratory, University of California, Berkeley, CA 94720, USA}%
	
	\author[0000-0002-6295-6978]{Michael J. MacDonald}
	\affil{Lawrence Livermore National Laboratory, 7000 East Avenue, Livermore, CA 94550, USA}
	
	\author[0000-0003-3894-5889]{Megan E. Eckart}
	\affil{Lawrence Livermore National Laboratory, 7000 East Avenue, Livermore, CA 94550, USA}
	
	\author{Richard L. Kelley}
	\affil{NASA Goddard Space Flight Center, 8800 Greenbelt Rd, Greenbelt, MD 20771, USA}
	
	\author[0000-0001-9464-4103]{Caroline A. Kilbourne}
	\affil{NASA Goddard Space Flight Center, 8800 Greenbelt Rd, Greenbelt, MD 20771, USA}
	
	\author[0000-0002-3331-7595]{Maurice A. Leutenegger}
	\affil{NASA Goddard Space Flight Center, 8800 Greenbelt Rd, Greenbelt, MD 20771, USA}
	
	\author[0000-0002-6374-1119]{F. Scott Porter}
	\affil{NASA Goddard Space Flight Center, 8800 Greenbelt Rd, Greenbelt, MD 20771, USA}
	
	\author[0000-0001-6338-9445]{Gregory V. Brown}
	\affil{Lawrence Livermore National Laboratory, 7000 East Avenue, Livermore, CA 94550, USA}

	%========= ABSTRACT ==================================================
	
	\begin{abstract}
		
		We report measurements of electron-impact excitation cross sections for the strong $K$-shell $n=2\rightarrow1$ transitions in \ion{S}{15} using the LLNL EBIT-I electron beam ion trap, two crystal spectrometers, and the EBIT Calorimeter Spectrometer. 
		The cross sections are determined by direct normalization to the well known cross sections of radiative electron capture, measured simultaneously. 
		Using contemporaneous polarization measurements with the two crystal spectrometers, whose dispersion planes are oriented parallel and perpendicular to the electron beam direction, the polarization of the direct excitation line emission is determined, and in turn the isotropic total cross sections are extracted. 
		We further experimentally investigate various line-formation mechanisms, finding that radiative cascades and collisional inner-shell ionization dominate the degree of linear polarization and total line-emission cross sections of the forbidden line $z$. 
	\end{abstract}
	
	
	%=====================================================================
	\section{Introduction}
	\label{intro}
	%=====================================================================

	High-resolution X-ray observations of various astrophysical objects by the \textit{Chandra} and \textit{XMM-Newton} satellites have provided unparalleled insights into their structure, composition, energy balance, mass-flow dynamics, density, and temperature distributions. 
	Among the most intense emission lines, originating from the transitions between ground level $1s^2 \,^1S_0$ and $1s2l$ levels in heliumlike ions dominate X-ray spectra observed from active galactic nuclei~\citep{porquet2000,bianchi2005}, supernova remnants~\citep{rasmussen2001,katsuda2012}, stellar coronae~\citep{audard2001,ness2003}, galaxy clusters~\citep{peterson2001,tamura2001,kaastra2001}, and solar flares~\citep{watanabe1995,murnion1996,sterling1997}. 
	The four most intense lines emanate from the upper level $1s2p \,^1P_1$, $1s2p \,^3P_2$, $1s2p \,^3P_1$, and $1s2s \,^3S_1$ to the ground state, known as resonance $w$, intercombination $x$, $y$ and forbidden $z$, respectively~\citep{gabriel1972}.
	The flux and energies of these transitions provide sensitive diagnostics of electron temperatures and densities, UV field strength, elemental abundances, ionization conditions, turbulent velocities, and opacities \citep{gabriel1969,freeman1971,doschek1971,blumenthal1972,mewe1978,kahn2001,leutenegger2006,porquet2010}.

	The high-resolution X-ray spectrum of the Perseus cluster, obtained using Hitomi's Soft X-ray Spectrometer (SXS) microcalorimeter has once again demonstrated the diagnostic power of heliumlike lines. 
	Their broadening and centroid shift was used to measure turbulent motion and shear at the center of the cluster~\citep{hitomi2016}. 
	However, detailed analysis of the Perseus spectrum uncovered additional challenges:
	the accuracy of atomic data employed in the standard plasma modeling codes, such as AtomDB/APEC~\citep{foster2012}, SPEX~\citep{kaastra1996}, and CHIANTI~\citep{delzanna2015}, did not meet the standard dictated by the \textit{Hitomi} SXS spectrum~\citep{hitomiatomic2018}.
	Furthermore, when determining how resonance scattering affected line $w$, its intensity relative to the optically-thin forbidden line $z$ was used.
	Large discrepancies in the intensity of line $z$ predicted by different plasma codes, however, affected the accuracy of the inferred amount of scattering~\citep{hitomi2018rs}.

	Line $z$ has a relatively complicated excitation structure, as its upper level can be populated in several ways. 
	In high-temperature plasmas, direct collisional excitation (CE) from the ground state represents only a small fraction of the total excitation function.
	Other channels include cascades from the $n$ = 2, 3, and from $n\geq$ 4 excited levels, inner-shell collisional ionization (CI) of the $1s^2 2s \,\,^2S_{1/2}$ ground state of Li-like ions, and radiative recombination from H-like ions. 
	Moreover, unresolved dielectronic recombination resonances~\citep{beiersdorfer1992} and charge exchange~\citep{wargelin2008} can also contribute to its apparent line strength.
	Therefore, accurate modeling of all these components is necessary to reliably predict the strength of line $z$.
	Lines $x$ and $y$ are also optically thin; however, they are weaker, and even at an energy resolution of 5 eV, are only marginally resolved from nearby lithiumlike dielectronic satellite lines.
	Furthermore, the heliumlike K$\beta$ line is also much weaker than the line $z$. 
	Thus, in sources where opacity affects the strength of the resonance line $w$, line $z$ often plays a significant role in determining the derived ion abundance and metallicity. 

	Future X-ray observatory missions, such as \textit{XRISM}~\citep{xrism2018} and \textit{Athena}~\citep{barret2016}, will also use high-resolution, high-throughput, wide-band X-ray microcalorimeter instruments, specifically Resolve, and X-IFU, respectively.
	These will observe strong He$\alpha$ line complexes from astrophysically abundant ions, and will determine the $w/z$ intensity ratio to an accuracy on the order of 5\%, given the strong heliumlike iron emission observed from Perseus by \textit{Hitomi}~\citep{hitomi2016}.
	Therefore, accurate atomic data become imperative for the reliable interpretation of future observations, lest the uncertainties in derived quantities will be dominated by atomic data, as opposed to limits on instrumentation or physical processes in the source. 

	With respect to the forbidden line $z$, little or no experimental data currently exist at the required level of accuracy. 
	Systematic measurements of all the processes contributing to the strength of line $z$ are also not available. 
	The data that do exist can only be used as a check at conditions near equilibrium~\citep{chantrenne1992,wong1995,bitter2008,beiersdorfer2009}. 
	However, in transient plasma conditions, such as in solar flares~\citep{mewe1980} and in supernova remnants~\citep{rasmussen2001}, where a sudden increase in the electron temperature can occur after a flare or shock, a significant fraction of Li-like ions can exist at high electron temperature together with the He-like ions.
	In such nonequilibrium plasma conditions, inner-shell collisional ionization of Li-like ions can significantly populate line $z$~\citep{decaux1995,decaux1997,bitter2008,porquet2010}. 
	Therefore, in order to model a wide range of sources and physical conditions, a systematic understanding of each population process has to be benchmarked using precise laboratory experiments. 
	
	An electron beam ion trap (EBIT) is an excellent tool with which to study such direct and indirect line-formation mechanisms, and to determine the relevant cross sections~\citep{beiersdorfer1990,wong1995,chen2002a,biedermann2002,brown2006,chen2006,chen2008,nakamura2008,ali2011,mahmood2012,shah2019,lindroth2020}. 
	The quasi-monoenergetic electron beam can be used to probe specific atomic processes, making it ideal for the benchmarking of plasma modeling codes.

	{
	One parameter that in some cases limits the accuracy of the EBIT measurement is the X-ray line polarization. 
	The unidirectional electron beam produces a preferred direction, and, in turn, a non-isotropic population distribution of magnetic sublevels~\citep{oppenheimer1927,percival1958,henderson1990}.
	The effects of polarization on measured line strengths are typically corrected using theoretical calculations.
	However, the theoretical modeling of line polarization is also not straightforward, as it must account for the device geometry, excitation mechanisms, collision direction, and photon decay paths~\citep{inal1987,reed1993,beiersdorfer1996,gu1999,robbins2004,takacs1996,hakel2007,hu2012,weber2015,shah2015,shah2016,shah2018,gall2020}.
	Consequently, polarization corrections contribute additional uncertainty, often comparable to the level of uncertainty required in astrophysical plasma models~\citep{hitomiatomic2018,hitomi2018rs}.
	Therefore, the line polarization must be independently benchmarked. 
	In contrast to previous experiments, we therefore measure the polarization simultaneously with the line-emission cross sections.
	}

    {Although the primary motivation for including polarization benchmarks is to increase the accuracy of EBIT cross-section measurements, polarization effects can also have direct relevance for astrophysical sources where non-isotropic electron-velocity distributions lead to the emission of anisotropic and polarized X-rays. 
    Such sources include solar flares~\citep{haug1972,haug1981,laming1990,laming1990a}, active galactic nuclei~\citep{nayakshin2007,dovciak2004}, and pulsars~\citep{weisskopf2006,kallman2004}.
	Measurements of polarization from these sources may reveal the presence and orientation of particle beams, magnetic fields, and distributions of nonthermal or suprathermal electrons, thereby providing information regarding plasma heating and confinement mechanisms.
	Besides, astrophysical spectropolarimetry is of particular interest, as it is often the only available technique for deriving information relating to the geometrical properties of angularly unresolved sources~\citep{krawczynski2011,soffitta2013,kallman2004}. 
	}

	{
	In this work, we simultaneously measure the electron-impact excitation cross sections and the polarization of strong heliumlike $1s2l$ lines of \ion{S}{15} using the EBIT. }
	Two orthogonal crystal spectrometers~\citep{beiersdorfer2016} are used to measure the X-ray polarization, and, simultaneously, the ECS microcalorimeter~\citep{porter2009} is used to obtain total cross sections. 
	Combining both allows us to measure the total effective line-emission cross sections to an accuracy of better than 10\%.
	In order to quantify the relative contribution of inner-shell ionization to the line strength of line $z$, we performed the measurements with two different relative fractions of Li- and He-like ions at multiple electron-impact energies. 
	We further compared the measured degree of linear polarization and total line-emission cross sections with relativistic distorted wave predictions.

\begin{figure}[t]
	\centering
	\includegraphics[clip=true,width=\columnwidth]{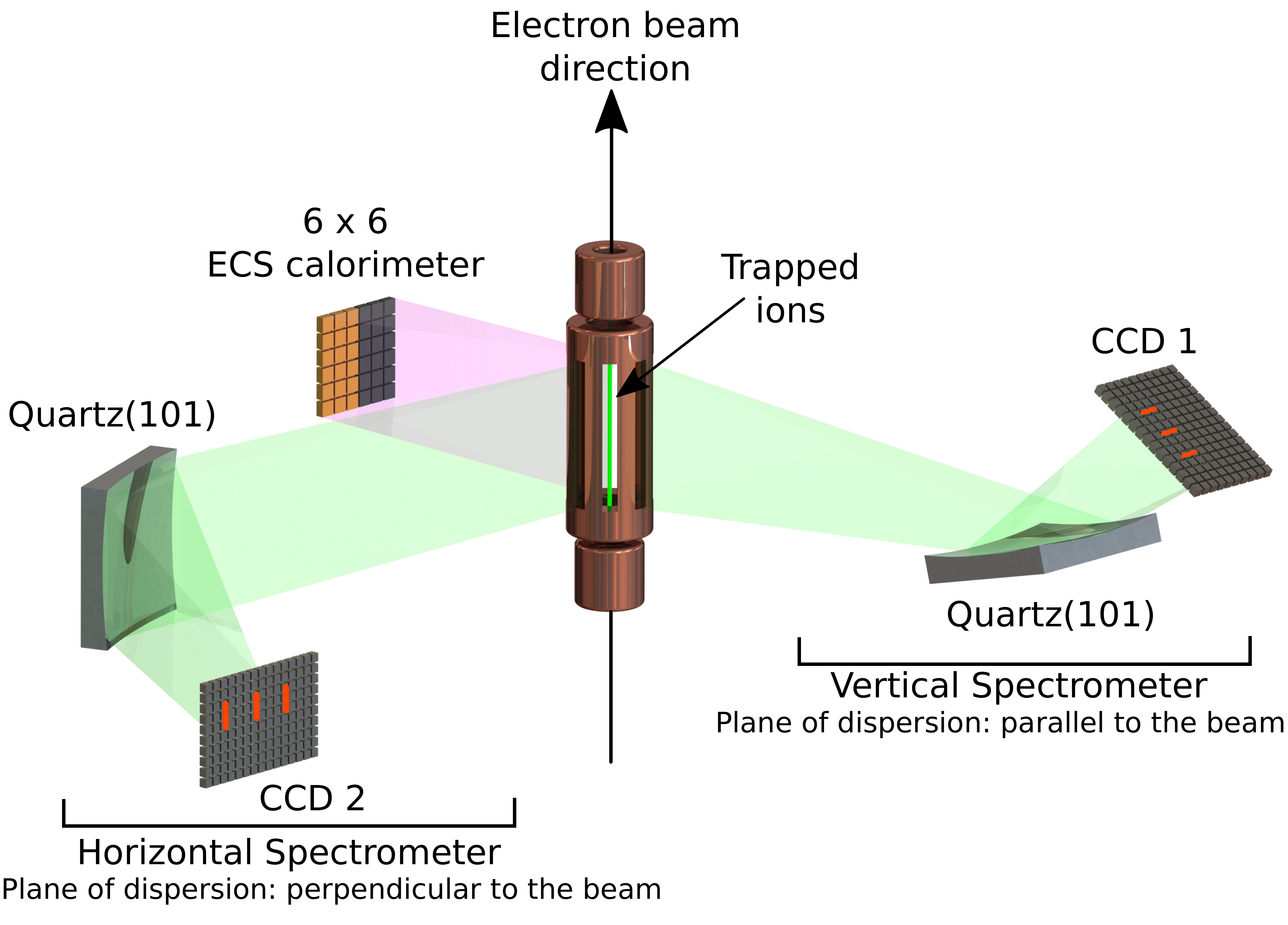} 
	\caption{
		Experimental setup layout: Ions are produced and trapped within a monoenergetic electron beam inside an EBIT. The X-rays, emitted from trapped ions, are simultaneously recorded by two EBIT high-resolution X-ray spectrometers (EBHiXs), whose dispersion planes are oriented parallel (vertical) and perpendicular (horizontal) to the electron beam direction, and by a wide-energy-band EBIT calorimeter spectrometer (ECS) array. 
	} 
	\label{fig:exp}
\end{figure}

	%=====================================================================
	\section{Experiment}
	\label{sec:exp}
	%=====================================================================

	The measurements were performed using Lawrence Livermore National Laboratory's electron beam ion trap, LLNL EBIT-I~\citep{levine1988}.
	In brief, EBIT-I produces an electron beam, originating from the hot surface of a cathode, which is accelerated toward a stack of cylindrical trap electrodes known as drift tubes. The beam is compressed to a diameter of approximately $60\,\mathrm{\mu m}$~\citep{levine1989, marrs1995} in the trap center by a 3-T magnetic fields generated by a pair of superconducting Helmholtz coils~\citep{herrmann1958}.
	Sulfur was introduced into the trap as SF$_6$, either via continuous injection with a two-stage differential pumping system for a lower average charge balance (1:1 He-like:Li-like fraction) or via  pulsed gas injection to reach a very high charge balance, dominated by He-like ions. 
	The electron beam dissociates SF$_6$ molecules, and ionizes sulfur atoms to high charge states via successive electron-impact ionization. 
	These highly charged S ions are axially confined by applying appropriate potentials to the drift tubes. 
	Simultaneously, radial confinement is provided by electrostatic attraction of the electron beam, as well as the flux freezing of the ions within the magnetic field.
	The trapped ions are also dumped periodically before beginning a new cycle of injection, charge breeding, and trapping. This prohibits the accumulation of high-$Z$ impurities, such as Ba and W, emerging from the electron gun~\citep{penetrante1991}. 

	X-rays emitted by trapped ions are observed at an angle of $90^\circ$ with respect to the electron beam propagation axis. 
	Owing to the unidirectional electron beam, the emitted X-rays are usually anisotropic and polarized. 
	{To quantify the X-ray polarization,} we utilized two imaging, high-resolution, spherically bent crystal spectrometers, dubbed as the EBIT high-resolution X-ray Spectrometer (EBHiX)~\citep{beiersdorfer2016a,beiersdorfer2016,hell2016}. 
	These are mounted parallel and perpendicular to the electron beam propagation axis, as depicted in Fig.~\ref{fig:exp}. 
	EBHiX adopts design parameters from spherically bent crystal spectrometers used for tokamaks~\citep{bitter2004}. It is designed such that the spatial focus coincides on the detector plane with the spectral focus of the Johann geometry, i.e., $D = R_c\sin\theta$, with a crystal radius of curvature $R_c$ and Bragg angle $\theta$. 
	The source position then follows from the focal relations of a spherical mirror. 
	The EBHiXs are designed around a nearly fixed nominal Bragg angle of 51.3$^\circ$ and a corresponding fixed source-to-crystal distance of 2.4~m, requiring a crystal radius of curvature of 67.2~cm.  
	We used two identical quartz (101) crystals, with a $2d$ spacing of 6.687~$\angstrom$ and a typical intrinsic resolving power, $E/\Delta E$, of 10,000~\citep{beiersdorfer2016}. 
	Both crystals are affixed to a small rotatable mount, which allows fine adjustment of the central Bragg angle to make small energy-range changes. 
	X-rays are recorded with nitrogen-cooled charge-coupled devices (CCDs) measuring $1300\times1340$ pixels, with each pixel having an area of $20~\mu\mathrm{m}~\times20~\mu\mathrm{m}$.
	{The vacuum of each EBHiX spectrometer is separated from EBIT-I by an aluminized polyimide window (0.1~$\mu$m~Al / 1.0~$\mu$m~polyimide). }

\begin{figure}%[ht]
		\centering
		\includegraphics[clip=true,width=\columnwidth]{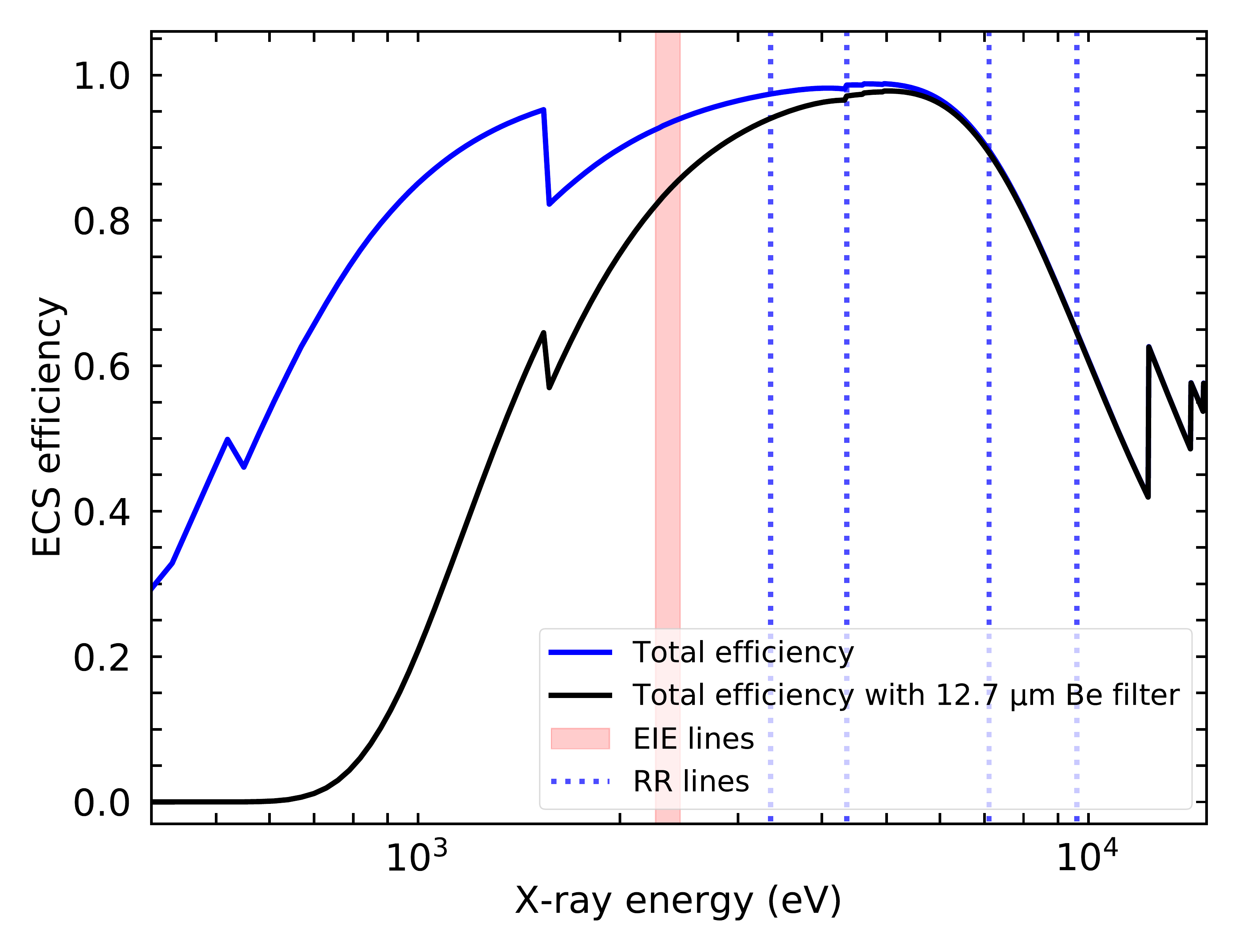} 
		\caption{Total EBIT calorimeter spectrometer (ECS) quantum efficiency, with and without $12.7~\mu\mathrm{m}$~Be filter. The energy positions of electron-impact excited (EIE) lines and radiative recombination (RR) lines of \ion{S}{15} are indicated by the vertical shaded area, and dotted lines, respectively.
		} 
		\label{fig:QE}
	\end{figure}

	To observe X-rays from a wide energy band, we used the EBIT Calorimeter Spectrometer (ECS), a $6\times6$ microcalorimeter array built at the NASA/Goddard Space Flight Center~\citep{porter2004,porter2008,porter2009}.
	{The ECS array, operated at $\sim60$~mK, consists of two types of pixels: 
	(a) low-energy pixels, optimized for 0.1 to 10 keV, having $8.59~\mu$m~thick HgTe absorbers with an area of $625 \mu\mathrm{m} \times 625 \mu\mathrm{m}$, giving a quantum efficiency of 95\% and an excellent energy resolution of $\sim$5 eV full-width-half-maximum (FWHM) at 6 keV, and 
	(b) high-energy pixels, covering a range from 0.5 to 100 keV, having an area of $625~\mu\mathrm{m} \times 500~\mu\mathrm{m}$ and $114~\mu$m~thick HgTe absorbers, which provide a much higher quantum efficiency, at high energies, of 32\% at 60 keV. }
	The ECS in its current configuration has four optical blocking filters in front of the array, totalling~0.15~$\mu$m~aluminum and 0.24~$\mu$m~polyimide. 
	In addition, there is a 0.05~$\mu$m~polyimide filter in the beam path to separate the ECS and EBIT-I vacua; and we also used a 12.7~$\mu$m~Be window to filter low-energy X-rays, particularly those from bright L-shell transitions, in order to keep the total count rate low. 
	{Figure~\ref{fig:QE} shows the total ECS quantum efficiency, including total filter transmission and absorber stopping power for low-energy pixels (see details in~\citet{thorn2008thesis,hell2017phd}). }
		
	Using both ECS and EBHiXs, we measured the \ion{S}{15} excitation cross sections at four different monoenergetic electron beam energies:~2.6,~3.6,~6.4, and~8.9~keV. 
	These energies were selected to be above the \ion{S}{15} K$\alpha$-excitation threshold, and to probe different compositions of the line formation contributions, i.e., (a) below the $n\geq3$ excitation and the Li-like inner-shell excitation thresholds (2.6\,keV), (b) to include cascades from $n\geq3$ and inner-shell ionization contributions (3.7\,keV), and (c) to enable a large relative H-like ion abundance in the trap (6.4\,keV). The crystal spectrometers were setup at the electron beam energy of 8.9 keV.
    All of these beam energies are free from dielectronic recombination (DR) and resonance excitation (RE) contributions. 
	Both the polarization and cross sections were measured at two different charge balances in the EBIT, in order to better diagnose the effects of inner-shell ionization. 
	The medium charge-balance (MC) state ($\sim50$\% He- and $\sim50$\% Li-like sulfur ions) was achieved via continuous gas injection at moderately high injection pressures of $2\times10^{-7}$\,Torr, providing a constant supply of low charge-state S ions in the trap. 
	By injecting a burst of gas with the pulsed gas injector once per EBIT cycle, immediately after the trap was dumped, we obtained a high charge-balance (HC) state ($\sim90$\% He-like ions). 
	The trapped ions were also dumped every 920~ms for 5~ms.

	%=====================================================================
	\section{Data Analysis}
	\label{analysis}
	%=====================================================================
	
	The line intensity observed by an ECSarray at $90^{\circ}$ due to electron-impact excitation (EIE) in an EBIT can be described as follows~\citep{wong1995}:
	\begin{equation}
	I_{90^{\circ}}^{\mathrm{EIE}} = \frac{j_e}{e} \,\,\, \sigma_{90^{\circ}}^{\mathrm{EIE}} \,\,\, n_{\mathrm{He}} \,\,\,  \Omega^{\mathrm{EIE}} \,\,\, \left( \eta\,T \right)^{\mathrm{EIE}},
	\label{eq:crossEIE}
	\end{equation}
	where $j_e$ is the effective current density, $e$ the electron charge, $n_{\mathrm{He}}$ the number density of heliumlike ions, $\Omega^{\mathrm{EIE}}$ the solid angle subtended by the calorimeter observing photons following EIE, $\left( \eta\,T \right)^{\mathrm{EIE}}$ the combined quantum efficiency and filter transmission at EIE energies, and
	$\sigma_{90^{\circ}}^{\mathrm{EIE}}$ is the line-emission cross section at $90^{\circ}$, which can be described as 
	\begin{equation}
	\sigma_{90^{\circ}} = \frac{ \sigma_{\mathrm{total}}}{4 \pi} \left(\frac{3}{3 \pm P}\right),
	\label{eq:cross90}
	\end{equation}
	where $P$ is the linear polarization of emitted X-rays. The negative sign should be used for an electric-dipole (E1) transitions and positive for a magnetic-dipole (M1) transitions.

	Likewise, the intensity of radiative recombination (RR) X-rays observed at $90^{\circ}$ can be written as
	\begin{equation}
	I_{90^{\circ}}^{\mathrm{RR}} = \frac{j_e}{e} \,\,\, \sigma_{90^{\circ}}^{\mathrm{RR}} \,\,\, n_{\mathrm{He}} \,\,\, \Omega^{\mathrm{RR}} \,\,\, \left( \eta\,T \right)^{\mathrm{RR}}.
	\label{eq:crossRR}
	\end{equation}

	The dependence on the effective current density and number of heliumlike ions can be eliminated if we simultaneously measure X-rays following EIE and RR~\citep{chantrenne1992},
	\begin{equation}
	\frac{I_{90^{\circ}}^{\mathrm{EIE}}}{I_{90^{\circ}}^{\mathrm{RR}}} = \frac{\sigma_{90^{\circ}}^{\mathrm{EIE}}}{\sigma_{90^{\circ}}^{\mathrm{RR}}} \,\, \left(\frac{\Omega^{\mathrm{EIE}}}{\Omega^{\mathrm{RR}}}\right) \,\, \frac{\left( \eta\,T \right)^{\mathrm{EIE}}}{\left( \eta\,T \right)^{\mathrm{RR}}}.
	\label{eq:ratio}
	\end{equation}
	By means of~Equations~\ref{eq:cross90} and~\ref{eq:ratio}, we can determine the total electron-impact cross section, relative to that of RR,~i.e.,
	\begin{equation}
	\sigma_{\mathrm{total}}^{\mathrm{EIE}} = 4 \pi \,\,\, \left(\frac{3 \pm P_{\mathrm{EIE}}}{3}\right)\,\,\,  \sigma_{90^{\circ}}^{\mathrm{RR}}  \left( \frac{I_{90^{\circ}}^{\mathrm{EIE}}}{I_{90^{\circ}}^{\mathrm{RR}}} \right)  \,\, \left(\frac{\Omega^{\mathrm{RR}}}{\Omega^{\mathrm{EIE}}}\right) \,\, \frac{\left( \eta\,T \right)^{\mathrm{RR}}}{\left( \eta\,T \right)^{\mathrm{EIE}}},
	\label{eq:totalcross}
	\end{equation}
	where $\Omega^{\mathrm{RR}}=\Omega^{\mathrm{EIE}}$, as the solid angle, is energy-independent, and we observe both the RR and EIE emission using the ECS low-energy array. 
	In this experiment, we measured $P_{\mathrm{EIE}}$ and ${I_{90^{\circ}}^{\mathrm{EIE}}} / {I_{90^{\circ}}^{\mathrm{RR}}}$, and used theoretical $\sigma_{90^{\circ}}^{\mathrm{RR}}$ in order to obtain the total line-emission cross sections.

	%=====================================================================
	\subsection{X-ray Line Polarization}
	\label{sec:pol}
	%=====================================================================

	According to \citet{beiersdorfer1996}, the X-ray line intensity observed using a crystal spectrometer is
	\begin{equation}
	I_{\mathrm{obs}} =  {R_\parallel I_\parallel + R_\perp I_\perp },
	\end{equation}
	where $R_\parallel$ and $R_\perp$ are the integrated crystal reflectivities for X-rays polarized parallel and perpendicular to the plane of dispersion, and the X-ray line polarization is
	\begin{equation}
	P = \frac {I_\parallel - I_\perp } {I_\parallel + I_\perp }.
	\label{eq:pol}
	\end{equation}

	As mentioned above, X-rays are recorded simultaneously using two crystal spectrometers, oriented in such a way that their planes of dispersion are parallel (vertical; V) and perpendicular (horizontal; H) to the electron beam propagation direction. 
	The line intensity observed by each spectrometer can then be written as
	\begin{equation}
	I_\mathrm{H} =  {R^\mathrm{H}_\parallel  I_\parallel + R^\mathrm{H}_\perp  I_\perp },
	\label{eq:hor}
	\end{equation}
	and
	\begin{equation}
	I_\mathrm{V} =  {R^\mathrm{V}_\parallel  I_\parallel + R^\mathrm{V}_\perp  I_\perp }.
	\label{eq:vert}
	\end{equation}
	The two crystals are identical, and the plane of dispersion of the vertical spectrometer is rotated by 90$^\circ$, as compared to that of its horizontal counterpart; thus we can write $R_\perp \equiv R^\mathrm{V}_\parallel = R^\mathrm{H}_\perp$ and $R_\parallel \equiv R^\mathrm{V}_\perp = R^\mathrm{H}_\parallel$, i.e., the perpendicular and parallel reflectivities of the vertical spectrometer are interchanged, compared to that of the horizontal, when assuming the same reference frame for the orientation of $I_\parallel$ and $I_\perp$. 
	Using this relation to solve Eqs.~\ref{eq:hor}~and~\ref{eq:vert} for $I_\parallel$ and $I_\perp$, and substituting them in Eq.~\ref{eq:pol}, we derive an expression for the degree of linear polarization of the X-rays, as a function of the observable intensities by the two crystal spectrometers: 
	\begin{equation}
	P =  \left( \frac {I_\mathrm{H} - I_\mathrm{V} }{I_\mathrm{H} + I_\mathrm{V} } \right) \,\, \left(   \frac {1+R} {1-R} \right) ,
	\label{eq:pol1}
	\end{equation}
	where the ratio $R = {R_\perp} / {R_\parallel}$ depends on the Bragg angle $\theta$, and can be estimated by $|\cos^2 2\theta|$ for a mosaic crystal, and  $|\cos 2\theta|$ for a perfect crystal. 
	Real crystals typically have $R$ values between these two limits~\citep{beiersdorfer1996}; however, we believe that the crystals used in our experiment are close to perfect. 
	We therefore calculated the $R$ values assuming perfect crystals, using \texttt{X0h}\footnote{\href{https://x-server.gmca.aps.anl.gov/x0h.html}{https://x-server.gmca.aps.anl.gov/x0h.html}}, which calculates the integrated reflectivities for quartz(101) in the energies of interest (2.4--2.5 keV) using data from~\citet{henke1993}. 
	These values were further verified using the X-ray Oriented Program~\citep[\textit{XOP};][]{XOP2011}~(see details in~\citet{macdonald2021}).
	
	\begin{figure}%[ht]
		\centering
		\includegraphics[clip=true,width=\columnwidth]{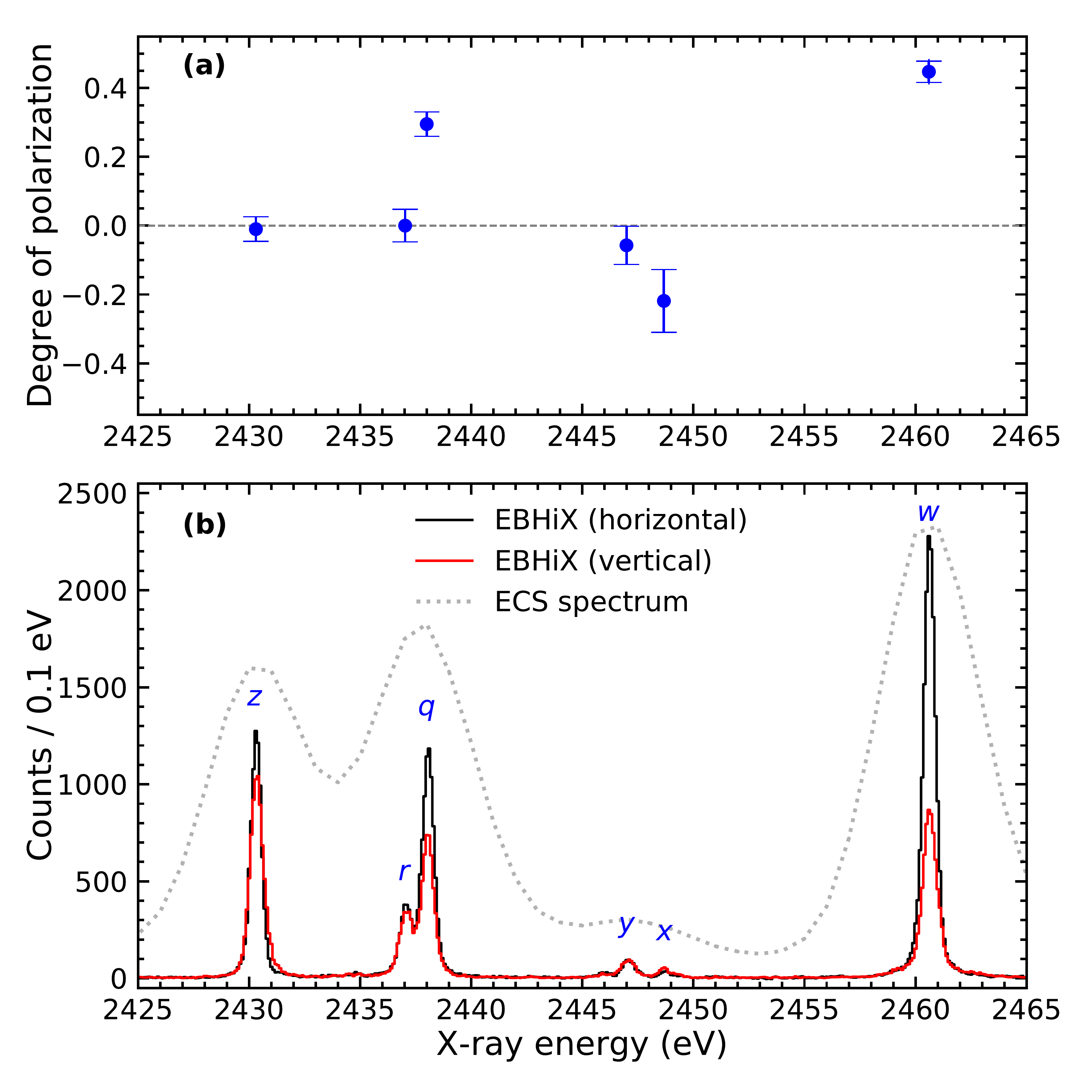} 
		\caption{
		(a) Measured degree of linear polarization for \ion{S}{15} $w, x, y$ and $z$ lines, and \ion{S}{14} $q$ line at a beam energy of 6.4 keV. 
		(b) Recorded X-ray emission spectra using two polarization-sensitive EBHiX crystal spectrometers: horizontal (black curve) and vertical (red curve). The non-polarization-sensitive ECS calorimeter spectrum, shown as a dotted curve, is also compared.
		} 
		\label{fig:pol65}
	\end{figure}

	The diffracted X-rays are recorded with CCD cameras, stabilized at $-110~^\circ$C, ensuring low thermal noise and dark current during their operations. 
	The CCD images were filtered for readout noise, cosmic rays, and other background events, following the method described by~\citet{pych2004}. 
	Subsequent to these corrections, the sum of all data is projected onto the axis corresponding to the dispersion direction.
	Several images, each with an hour's exposure, were obtained at each electron beam energy. 
	The spectra from all exposures were added together in order to improve the signal-to-noise ratio. 

	Figure~\ref{fig:pol65} shows an example of a summed spectrum at 6.5 keV of beam energy, with a medium charge-balance setting. 
	The wavelength calibrations for both spectrometers were performed using the theoretical value of line $w$ from \citet{drake1986}, and the experimentally measured value of line $z$ of \ion{S}{15}~\citep[][relative to \citet{drake1986}]{hell2016}. 
	The observed spectral peaks are then fitted using multiple Voigt functions, i.e., a convolution of Lorentz and Gaussian functions. 
	The forbidden line $z$ at 2430~eV has a very narrow natural line width ($\sim 9 \cdot 10^{-10}$ eV, see~\cite{crespo2006}) compared to our instrumental resolution.
	This line allows us to determine the spectral-line broadening due to the Doppler motion of the trapped ions and due to any contribution of an intrinsic spectrometer line-spread function. 
	{Based on the axial trap depth ($V_0\sim50$~V), the expected initial ion temperature ($T_i\sim$140~eV), and the crystal spectrometer line shape~\citep{macdonald2021}, the observed Gaussian width, which is 0.47-eV FWHM at 2430~eV~($T_i \sim$200~eV), should be dominated by the Doppler width, while the dominant contribution to the Lorentzian width is due to the instrument, with a negligible part due to the natural linewidth.
	Thus, in our fitting procedure, the Gaussian width obtained from line $z$ is shared between all peaks, while Lorentzian widths, representing the sum of instrument and natural line widths, are allowed to vary, as are the peak centroids and amplitudes.}

	Any photon-energy independent differences between the two crystal spectra that may arise from differences in their geometry and efficiency can be corrected using an unpolarized line for cross-normalization. X-ray lines with total angular momentum in the upper level of $J\leq{1/2}$ are unpolarized~\citep{balashovbook}.    
	In this experiment, we used an unpolarized line, $r$ (excited state $[1s\,2s\,2p_{1/2}]_{J={1/2}}$), from lithiumlike \ion{S}{14} at 2437~eV to normalize the relative efficiencies of the vertical and horizontal spectrometers. 

	Finally, the normalized intensities and relative reflectivities, $R$, obtained from \textit{XOP} are used in Eq.~\ref{eq:pol1} to determine the degree of linear polarization of an X-ray line following the EIE. 
	The uncertainty in the measured degree of polarization includes contributions from statistical fitting error of the line intensities, background removal, crystal reflectivities, and the relative efficiency normalization between the two crystal spectrometers inferred from the lithiumlike \ion{S}{14} line $r$. %(1.1$\pm$0.05)
	The results obtained for $P_{\mathrm{EIE}}$ at a beam energy of 6.4 keV are shown in the top panel of Fig.~\ref{fig:pol65}. 
	We followed a similar procedure for data sets with electron beam energies of 2.6, 3.6, and 8.9~keV. 
	Given that lines $w$ and $z$ are well isolated, we checked the validity of our fitted line intensities by integrating counts over their respective energy ranges, obtaining degree-of-polarization results within the measurement uncertainty. 
	%=====================================================================
	\subsubsection{RR Cross Sections}
	\label{sec:rr-theort}
	%=====================================================================
				
	\begin{figure}
		\centering
		\includegraphics[clip=true,width=\columnwidth]{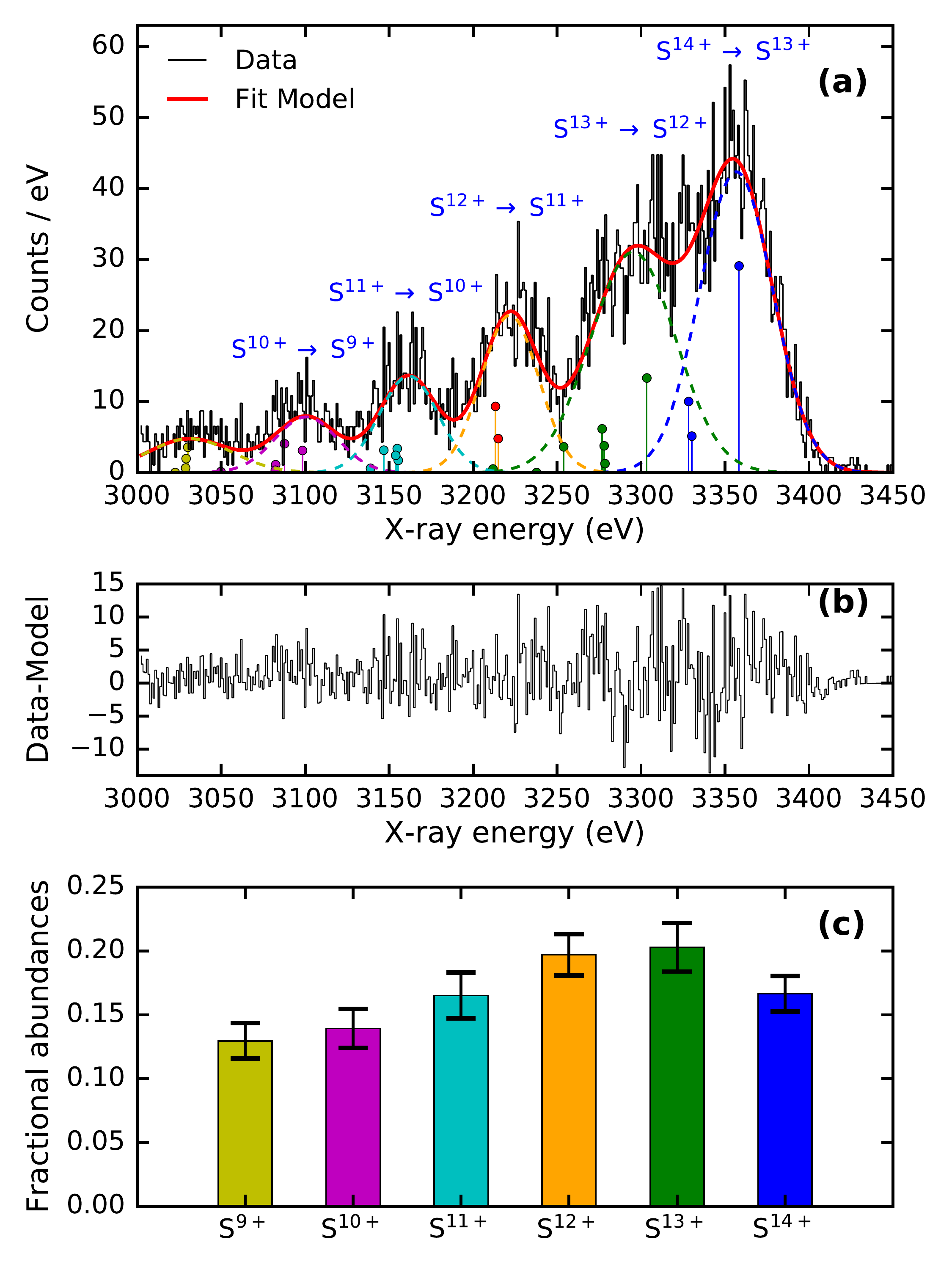} 
		\caption{
		Top panel (a): S RR spectrum and spectral fit to the measured data. The vertical stems indicate the position and relative intensity of individual RR lines from different charge states. 
		Middle panel (b): fit residuals. 
		Bottom panel (c): relative charge balance inferred from the RR analysis of medium charge balance measurement taken at 2.6 keV beam energy. 
		} 
		\label{fig:rr26}
	\end{figure}

	\begin{table*}[ht]
		\centering
		%\scriptsize
		\caption{\label{tab:RR} 
			Tabulated energy-dependent differential RR cross sections (in units of $10^{-24} \mathrm{cm}^2$) using FAC~\citep{gu2008}. They are expressed as coefficients from a fifth-order polynomial fit, $\sigma_{RR} = \sum_{i=0}^5 a_i E^{-i}\,$\citep{chen2005} with E in keV and valid in the range 1.5--10\,keV.}
		\begin{tabular}{ccccccccccc}
			\hline \hline			
			\multicolumn{3}{c}{Radiative recombination into} & Binding energy & \multicolumn{6}{c}{Fitting Coefficients} \\
			\cline{1-3}\cline{5-11}
			ion & $n$   & State & (eV)  && $a_0$   & $a_1$   & $a_2$   & $a_3$   & $a_4$   & $a_5$ \\
			\hline 
			H-like \ion{S}{16} 	&	1	&	$\left[	1s_{1/2}\right]_{1/2}	$	&	3494.2	&&	 $-$0.394 	&	9.176	&	199.311	&	 $-$412.093 	&	463.886	&	$-$213.774  	\\		
He-like \ion{S}{15} 	&	1	&	$\left[	1s_{1/2}\right]_{0}	$	&	3221.9	&&	 $-$0.188 	&	4.415	&	95.322	&	 $-$196.599 	&	220.678	&	$-$101.434  	\\		
H-like \ion{S}{16} 	&	2	&	$\left[	2s_{1/2}\right]_{1/2}	$	&	874.4	&&	 $-$0.042 	&	0.800	&	30.393	&	 $-$60.336 	&	67.080	&	$-$31.050  	\\		
H-like \ion{S}{16} 	&	2	&	$\left[	2p_{1/2}\right]_{1/2}	$	&	874.5	&&	0.005	&	 $-$0.182 	&	2.416	&	7.469	&	 $-$12.594 	&	6.614	\\		
H-like \ion{S}{16} 	&	2	&	$\left[	2p_{3/2}\right]_{3/2}	$	&	871.5	&&	0.010	&	 $-$0.353 	&	4.564	&	15.153	&	 $-$25.245 	&	13.210	\\		
He-like \ion{S}{15} 	&	2	&	$\left[	1s2s_{1/2}\right]_{0}	$	&	774.3	&&	 $-$0.011 	&	0.225	&	6.910	&	 $-$14.167 	&	15.998	&	$-$7.470  	\\		
He-like \ion{S}{15} 	&	2	&	$\left[	1s2s_{1/2}\right]_{1}	$	&	792.8	&&	 $-$0.023 	&	0.410	&	20.396	&	 $-$39.218 	&	43.028	&	$-$19.821  	\\		
He-like \ion{S}{15} 	&	2	&	$\left[	1s2p_{1/2}\right]_{0}	$	&	776.0	&&	0.001	&	0.001	&	0.438	&	1.624	&	 $-$2.532 	&	1.279	\\		
He-like \ion{S}{15} 	&	2	&	$\left[	1s2p_{1/2}\right]_{1}	$	&	775.6	&&	0.003	&	 $-$0.096 	&	1.295	&	4.890	&	 $-$7.606 	&	3.843	\\		
He-like \ion{S}{15} 	&	2	&	$\left[	1s2p_{3/2}\right]_{1}	$	&	761.5	&&	0.003	&	 $-$0.093 	&	1.243	&	4.699	&	 $-$7.416 	&	3.771	\\		
He-like \ion{S}{15} 	&	2	&	$\left[	1s2p_{3/2}\right]_{2}	$	&	774.0	&&	0.004	&	 $-$0.153 	&	2.033	&	8.331	&	 $-$12.868 	&	6.501	\\		
Li-like \ion{S}{14} 	&	2	&	$\left[	2s_{1/2}\right] _{ 1/2}	$	&	706.7	&&	 $-$0.033 	&	0.605	&	26.369	&	 $-$55.433 	&	64.199	&	$-$30.534  	\\		
Li-like \ion{S}{14} 	&	2	&	$\left[	2p_{1/2}\right] _{ 1/2}	$	&	678.6	&&	0.005	&	 $-$0.165 	&	1.954	&	5.216	&	 $-$8.670 	&	4.481	\\		
Li-like \ion{S}{14} 	&	2	&	$\left[	2p_{3/2}\right] _{ 3/2}	$	&	676.7	&&	0.010	&	 $-$0.320 	&	3.692	&	10.650	&	 $-$17.470 	&	9.003	\\		
Be-like \ion{S}{13} 	&	2	&	$\left[	2s_{1/2}2s_{1/2}\right] _{ 0}	$	&	651.8	&&	 $-$0.014 	&	0.252	&	12.063	&	 $-$24.985 	&	28.614	&	$-$13.500  	\\		
Be-like \ion{S}{13} 	&	2	&	$\left[	2s_{1/2}2p_{1/2}\right] _{ 0}	$	&	626.9	&&	0.001	&	 $-$0.039 	&	0.464	&	1.306	&	 $-$2.077 	&	1.055	\\		
Be-like \ion{S}{13} 	&	2	&	$\left[	2s_{1/2}2p_{1/2}\right] _{ 1}	$	&	626.4	&&	0.004	&	 $-$0.117 	&	1.376	&	3.911	&	 $-$6.203 	&	3.148	\\		
Be-like \ion{S}{13} 	&	2	&	$\left[	2s_{1/2}2p_{3/2}\right] _{ 2}	$	&	625.2	&&	0.006	&	 $-$0.192 	&	2.208	&	6.582	&	 $-$10.346 	&	5.235	\\		
Be-like \ion{S}{13} 	&	2	&	$\left[	2s_{1/2}2p_{3/2}\right] _{ 1}	$	&	602.3	&&	0.004	&	 $-$0.118 	&	1.363	&	3.710	&	 $-$6.049 	&	3.101	\\		
Be-like \ion{S}{13} 	&	2	&	$\left[	2p_{1/2}2p_{1/2}\right] _{ 0}	$	&	586.2	&&	 $-$0.000 	&	0.000	&	0.003	&	 $-$0.007 	&	0.007	&	$-$0.003  	\\		
Be-like \ion{S}{13} 	&	2	&	$\left[	2p_{3/2}2p_{3/2}\right] _{ 0}	$	&	560.2	&&	 $-$0.000 	&	0.003	&	0.474	&	 $-$0.950 	&	1.038	&	$-$0.474  	\\		
B-like \ion{S}{12} 	&	2	&	$\left[	2p_{1/2}\right] _{ 1/2}	$	&	563.3	&&	0.005	&	 $-$0.149 	&	1.734	&	4.918	&	 $-$7.590 	&	3.791	\\		
B-like \ion{S}{12} 	&	2	&	$\left[	2p_{3/2}\right] _{ 3/2}	$	&	561.7	&&	0.008	&	 $-$0.277 	&	3.192	&	10.342	&	 $-$15.826 	&	7.943	\\		
C-like \ion{S}{11} 	&	2	&	$\left[	2p_{1/2}2p_{1/2}\right] _{ 0}	$	&	503.8	&&	0.001	&	 $-$0.048 	&	0.569	&	1.957	&	 $-$2.957 	&	1.485	\\		
C-like \ion{S}{11} 	&	2	&	$\left[	2p_{1/2}2p_{3/2}\right] _{ 1}	$	&	503.2	&&	0.003	&	 $-$0.095 	&	1.091	&	3.912	&	 $-$5.836 	&	2.912	\\		
C-like \ion{S}{11} 	&	2	&	$\left[	2p_{1/2}2p_{3/2}\right] _{ 2}	$	&	495.2	&&	0.003	&	 $-$0.090 	&	1.035	&	3.621	&	 $-$5.471 	&	2.743	\\		
C-like \ion{S}{11} 	&	2	&	$\left[	2p_{3/2}2p_{3/2}\right] _{ 2}	$	&	502.3	&&	0.002	&	 $-$0.069 	&	0.786	&	2.812	&	 $-$4.200 	&	2.097	\\		
C-like \ion{S}{11} 	&	2	&	$\left[	2p_{3/2}2p_{3/2}\right] _{ 0}	$	&	487.3	&&	0.001	&	 $-$0.017 	&	0.201	&	0.668	&	 $-$1.035 	&	0.525	\\		
N-like \ion{S}{10} 	&	2	&	$\left[	2p_{3/2}\right] _{ 3/2}	$	&	436.1	&&	0.003	&	 $-$0.106 	&	1.214	&	5.065	&	 $-$7.541 	&	3.795	\\		
N-like \ion{S}{10} 	&	2	&	$\left[	2p_{1/2}2p_{3/2}2p_{3/2}\right] _{ 3/2}	$	&	446.8	&&	0.002	&	 $-$0.079 	&	0.908	&	3.937	&	 $-$5.770 	&	2.888	\\		
N-like \ion{S}{10} 	&	2	&	$\left[	2p_{1/2}2p_{3/2}2p_{3/2}\right] _{ 1/2}	$	&	430.8	&&	0.001	&	 $-$0.029 	&	0.342	&	1.344	&	 $-$2.038 	&	1.032	\\		
N-like \ion{S}{10} 	&	2	&	$\left[	2p_{3/2}\right] _{ 3/2}	$	&	430.6	&&	0.000	&	 $-$0.009 	&	0.102	&	0.418	&	 $-$0.627 	&	0.317	\\		
N-like \ion{S}{10} 	&	2	&	$\left[	2s_{1/2}2p_{3/2}2p_{3/2}\right] _{ 1/2}	$	&	398.2	&&	 $-$0.000 	&	0.001	&	0.118	&	 $-$0.216 	&	0.211	&	$-$0.087  	\\		
O-like \ion{S}{9} 	&	2	&	$\left[	2p_{3/2}2p_{3/2}\right] _{ 2}	$	&	378.5	&&	0.003	&	 $-$0.091 	&	1.027	&	4.545	&	 $-$6.651 	&	3.306	\\		
O-like \ion{S}{9} 	&	2	&	$\left[	2p_{3/2}2p_{3/2}\right] _{ 0}	$	&	377.2	&&	0.000	&	 $-$0.016 	&	0.183	&	0.838	&	 $-$1.225 	&	0.609	\\		
O-like \ion{S}{9} 	&	2	&	$\left[	2p_{1/2}2p_{3/2}\right]_{ 1}	$	&	377.5	&&	0.001	&	 $-$0.050 	&	0.563	&	2.546	&	 $-$3.721 	&	1.851	\\		
O-like \ion{S}{9} 	&	2	&	$\left[	2p_{1/2}2p_{3/2}\right] _{ 2}	$	&	371.0	&&	0.000	&	 $-$0.000 	&	0.002	&	0.006	&	 $-$0.010 	&	0.005	\\		
F-like \ion{S}{8} 	&	2	&	$\left[	2p_{3/2}\right] _{ 3/2}	$	&	327.4	&&	0.003	&	 $-$0.088 	&	0.951	&	4.518	&	 $-$6.403 	&	3.112	\\		
F-like \ion{S}{8} 	&	2	&	$\left[	2p_{1/2}\right] _{ 1/2}	$	&	326.1	&&	0.000	&	 $-$0.013 	&	0.138	&	0.662	&	 $-$0.939 	&	0.456	\\		
Ne-like \ion{S}{7} 	&	2	&	$\left[	2p_{3/2}\right] _{ 0}	$	&	279.2	&&	0.002	&	 $-$0.048 	&	 $-$0.048 	&	2.527	&	 $-$3.452 	&	1.628	\\		
			\hline \hline
		\end{tabular}
	\end{table*}

	We used the Flexible Atomic Code ~\citep[FAC,][]{gu2008}\footnote{\href{https://github.com/flexible-atomic-code/fac}{https://github.com/flexible-atomic-code/fac}} to calculate the radiative recombination (RR) cross sections. 
	Similarly to~\citet{zhang1998}, FAC uses a fully relativistic distorted wave (DW) method to calculate non-resonant photoionization (PI) and RR processes.  
	RR cross sections are obtained via detailed balance from PI cross sections. 
	With respect to highly charged ions, electron correlation effects are less important for PI and RR processes. Thus, RR cross sections can be calculated with an accuracy of 5\% or even better~\citep{saloman1988,scofield1989,brown2006,chen2008}. 
	
	We also used FAC to calculate the polarization of X-rays emitted following the RR process, as photons are observed at 90$^\circ$ relative to the direction of electron impact in our experiment.
	To provide an energy-dependent RR cross sections for \ion{S}{6}~--\ion{}{16} ions, we fit a fifth-order polynomial in the variable $(1/E)$, with $E$ given in keV, to the calculated cross sections \citep{chen2005}. 
	The fitting parameters for the obtained RR cross sections of \ion{S}{16}~--\ion{}{7} ions in the range of 1.5--10 keV are listed in Tab.~\ref{tab:RR}.

	%=====================================================================
	\subsubsection{RR Line Intensity}
	\label{sec:rr-exp}
	%=====================================================================
	
	%
	The wide bandpass of the ECS microcalorimeter allows us to simultaneously measure X-rays due to RR and EIE. 
	The energy scale of the ECS was calibrated using several measurements of $K$- and $L$-shell EIE lines from different hydrogenlike and heliumlike O, Ne, Si, S, Ar, Fe, Ni, Ge, Kr, and Ne-like Ba ions.  
	Each low-energy and high-energy pixel was individually calibrated, using a fifth-order polynomial function. 
	{The spectrum is then normalized by the ECS efficiency curve including a $12.7~\mu\mathrm{m}$ Be filter, as discussed in Sec.~\ref{sec:exp} and shown in~Fig.~\ref{fig:QE}. 
	Further details regarding the ECS efficiency and associated uncertainties can be found in~\citet{thorn2008thesis} and \citet{hell2017phd}. }
	Figure~\ref{fig:rr26}(b) shows an example of an RR spectrum measured at an electron beam energy of 2.6 keV, with medium charge-balance settings. 

\begin{table}[b]
	\centering
	\caption{Ion fraction of \ion{S}{14} and \ion{O}{7} relative to that of \ion{S}{15}. }
	\begin{tabular}{cccc}
		\hline\hline
		Charge balance & MC (\ion{S}{15}$\sim$1.0)  & \multicolumn{2}{c}{HC (\ion{S}{15}$\sim$1.0) } \\
		\cline{3-4}
		Beam energy (eV) & \ion{S}{14} & \ion{S}{14} & \ion{O}{7} \\
		\hline
		2650  & 1.05  & 0.08  & 0.21\\%(3) \\
		3650  & 1.04  & 0.10  & 0.38\\%(5) \\
		6400  & 0.80  & 0.24  & 0.16\\%(2) \\
		8900  & 1.10  &       &  \\
		\hline\hline
	\end{tabular}%
	\label{tab:cs}%
\end{table}%

	The energies of X-ray peaks from RR are given by the sum of electron beam energy and the ionization potential of the state being recombined into. 
	We use this relation in our fitting procedure, as a first step, to obtain the precise energy of the RR lines from each charge state, as well as the effective electron beam energy, accounting for the negative space charge of the electron beam as well as ion compensation. 
	In this step, we fix the difference between the line centroids as the ionization potential difference between different charge states, as obtained from NIST~\citep{NIST_ASD}.
	We also share the Gaussian width between all peaks, as this is dominated by a convolution of the narrow Gaussian electron beam energy distribution in the EBIT and the Gaussian instrumental resolution. 
	Subtracting the instrument resolution in quadrature, we can thus determine the energy spread of the electron beam, e.g., $49\pm1$ eV FWHM at an electron beam energy of 2.6 keV with medium charge-balance settings. 
	This width is enough to allow us to distinguish the RR features from different charge states of sulfur ions. However, it is very large compared to the fine-structure components of an individual RR peak (see vertical stems in Fig.~\ref{fig:rr26}(b) and~Tab.~\ref{tab:RR}). 
	Hence, we produced synthetic RR spectra for each charge state based on the theoretical RR energies and cross sections at 90$^{\circ}$, and convolved them with the electron beam energy spread obtained in the previous step. 
	The convolved synthetic spectrum for each charge state is then fitted to the data to obtain centroids and widths. 
	These parameters, accounting for the fine structure components, are then used to fit the experimental RR spectrum to extract RR intensities.

	Using Eq.~\ref{eq:crossRR}, we inferred the relative charge balance of the trapped ions by taking the ratios of the measured RR intensities and cross sections. 
	The derived relative ion abundances are shown in Fig.~\ref{fig:rr26}(a) and in Tab.~\ref{tab:cs}.
	Note that the relative charge balance between heliumlike to lithiumlike S was also confirmed using the $w/q$ line ratios observed using the two crystal spectrometers.

	%=====================================================================
	\subsubsection{EIE Line Intensity}
	\label{exp:xsec}
	%=====================================================================
		
	\begin{figure}%[ht]
		\centering
		\includegraphics[clip=true,width=\columnwidth]{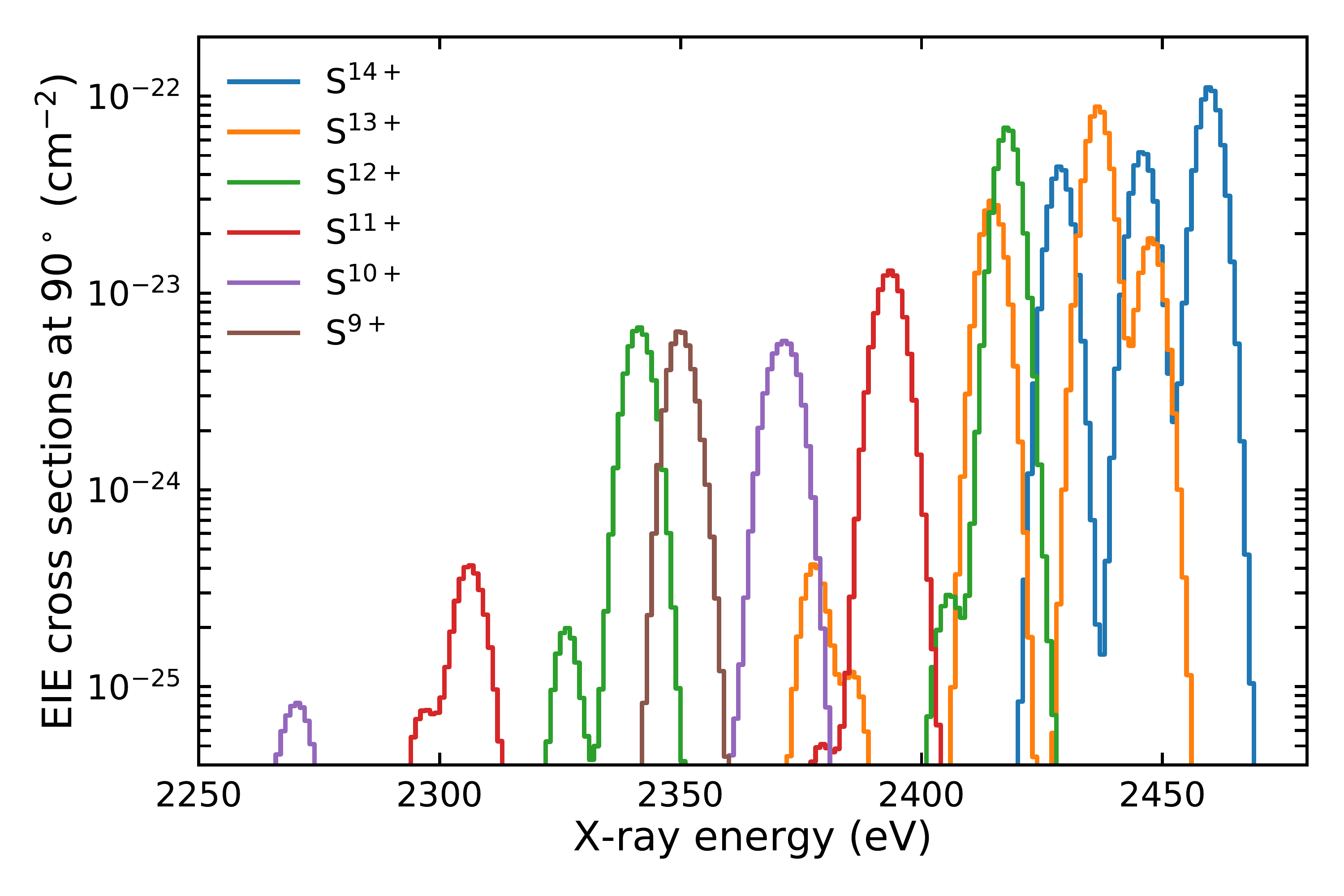} 
		\caption{
			Synthetic EIE spectrum, corrected for 90$^\circ$ observation at a 2.6 keV beam energy. 
		} 
		\label{fig:thEIE}
	\end{figure}

	The extraction of EIE line intensities is more complicated when more than one charge state is involved in the analysis and lines are blended. 
	To account for these, we used a collisional-radiative model, coupled with measured values where available, to determine the line positions and intensities. 
	
	%=====================================================================
	\paragraph{Synthetic EIE Spectrum}
	\label{th-EIE}
	%=====================================================================

	We used the latest version of FAC (v1.1.5) to compute the electronic structure of a given ion. 
	The fully relativistic distorted wave method was used to calculate the EIE cross sections. 
	For our calculations, the ground state configurations are added as $1s^2 \, (2l)^{(k-2)}$ and the excited state configurations as $1s \, (2l)^{(k-1)}$ and $1s \, (2l)^{(k-2)} \, (nl)$, where $k$ represents the number of electrons in \ion{S}{15}~--\ion{}{10} ions. 
	We considered principal quantum numbers up to $n=16$, ogether with all their possible angular momentum $l$ states, and included full mixing between all states in our calculations.
	
	The directional collisions between electrons and ions inside an EBIT usually lead to nonstatistical populations of magnetic sublevels of the upper state, which produce anisotropic and polarized X-rays~\citep{beiersdorfer1996}.  
	Therefore, EIE cross sections between the magnetic sublevels of upper and lower states were computed using FAC. 
	The X-ray polarization is then obtained from the calculated magnetic-sublevel cross sections $\sigma_{m_j}$, and intrinsic anisotropic factors, $\alpha_2$, for each line from \ion{S}{10}~--\ion{}{15} ions. 
	We also accounted for depolarization due to radiative cascades, and the transverse component of electron energy due to the cyclotron motion of electrons inside the electron beam~\citep{gu1999}. 
	Using the optical theory of electron beam propagation by~\citet{herrmann1958}, we estimated the transversal energy component to be $\sim$180~eV. 
	The optical theory predictions are generally found to be in good agreement with the laboratory measurements~\citep{beiersdorfer2001,shah2018}. 

	The produced atomic data, including line energies, transition rates, autoionization rates, and magnetic-sublevel-resolved EIE collision strengths, were then fed into the collisional-radiative model of FAC~\citep{gu2008}. 
	This solves a system of coupled rate equations to obtain level populations for specified input experimental conditions, such as the electron beam energy and density. 
	The resulting level populations are then used to produce a synthetic EIE spectrum at each electron beam energy. 
	An example is shown in Fig.~\ref{fig:thEIE}, where each line is convolved with the ECS calorimeter resolution. 
	The synthetic spectrum allows us to account for all possible weak transitions, which have a non-negligible contribution to the measured line intensity, particularly in lower charge state Be-~, B-~, and C-like S ions. 
	Furthermore, the relative intensities also take into account the inferred fractional abundances from the RR analysis. 
	
	%=====================================================================
	\paragraph{Experimental EIE Spectrum}
	\label{exp-EIE}
	%=====================================================================

	\begin{figure}
		\centering
		\includegraphics[clip=true,width=\columnwidth]{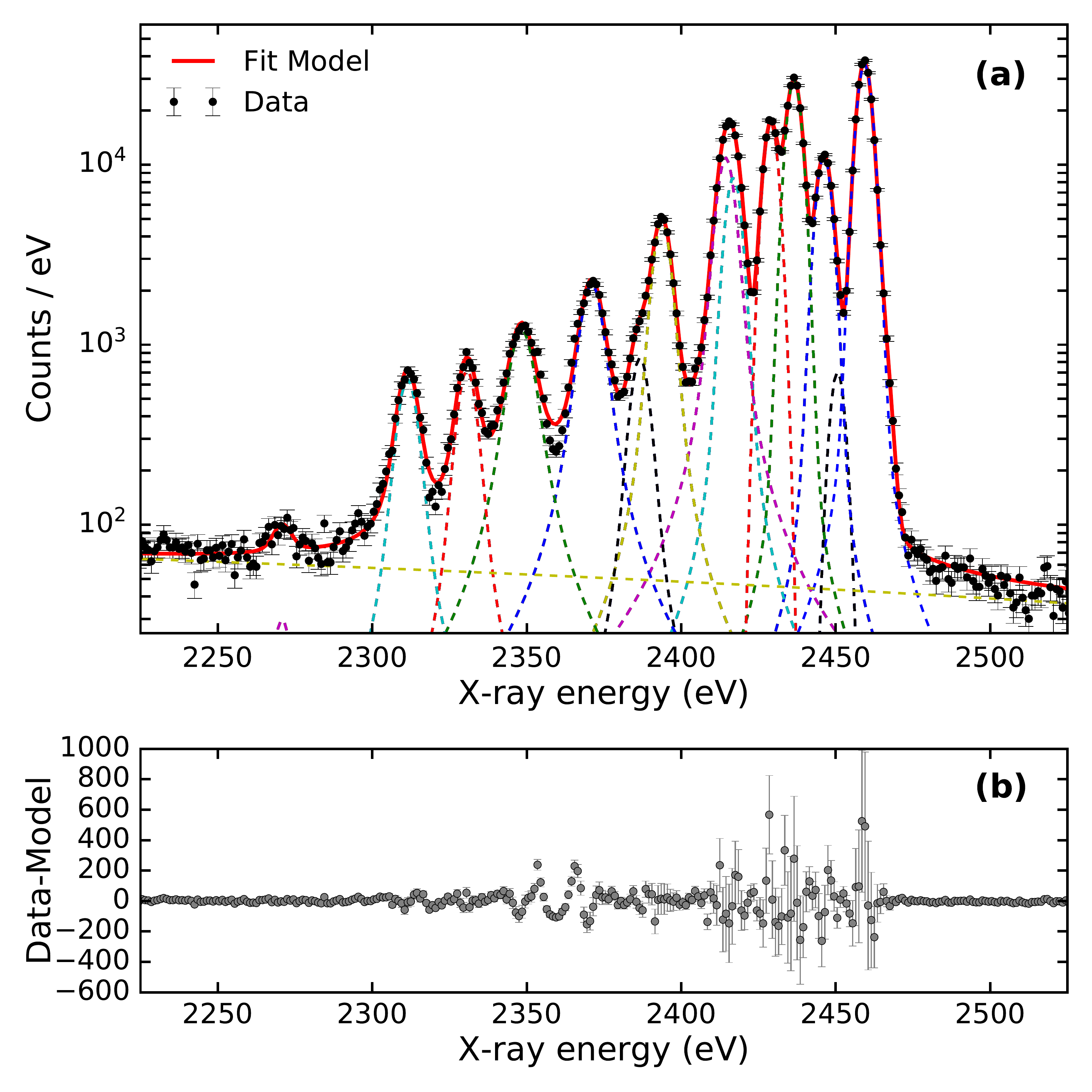} 
		\caption{
			S $K$-shell EIE spectrum at an electron-impact energy of 2.6 keV and best-fit model.
		} 
		\label{fig:de26}
	\end{figure}

	\begin{table*}
		\centering
		\caption{Measured degree of linear polarization and total electron-impact excitation cross sections in $10^{-21} \mathrm{cm}^2$. Terms MC and HC stand for medium charge balance and high charge balance, respectively.}
			\begin{tabular*}{\textwidth}{c @{\extracolsep{\fill}}cccc}
				\hline	\hline																
				Beam energy (eV)	&	$w$ polarization (MC)			&	$w$ polarization (HC)			&	$w$ total cross sections (MC)			&	$w$ total cross sections (HC)			\\
				\hline																	
				2650	&	0.56	$\pm$	0.03	&	0.57	$\pm$	0.03	&	2.65	$\pm$	0.18	&	2.57	$\pm$	0.19	\\
				3650	&	0.53	$\pm$	0.03	&	0.53	$\pm$	0.03	&	3.48	$\pm$	0.24	&	3.70	$\pm$	0.40	\\
				6400	&	0.45	$\pm$	0.03	&	0.50	$\pm$	0.02	&	3.08	$\pm$	0.48	&	3.45	$\pm$	0.26	\\
				8900	&	0.31	$\pm$	0.06	&				&	4.06	$\pm$	0.55	&				\\
				\hline																	
				&	$z$ polarization (MC)			&	$z$ polarization (HC)			&	$z$ total cross sections (MC)			&	$z$ total cross sections (HC)			\\
				\hline																	
				2650	&	$-0.19$	$\pm$	0.04	&	$-0.19$	$\pm$	0.02	&	1.33	$\pm$	0.09	&	1.35	$\pm$	0.10	\\
				3650	&	$-0.09$	$\pm$	0.04	&	$-0.20$	$\pm$	0.02	&	2.01	$\pm$	0.14	&	1.54	$\pm$	0.17	\\
				6400	&	$-0.01$	$\pm$	0.04	&	$-0.13$	$\pm$	0.03	&	2.04	$\pm$	0.18	&	0.94	$\pm$	0.07	\\
				8900	&	0.00	$\pm$	0.05	&				&	2.60	$\pm$	0.36	&				\\
				\hline	\hline																
			\end{tabular*}%
		\label{tab:results}%
	\end{table*}%

	We first initialize the EIE spectrum fit using the initial values of line positions and relative intensities from the synthetic spectrum. 
	In our fitting procedure, we use a single value of the Gaussian width for the Voigt profiles, representing the quadrature sum of the core of the calorimeter line spread function and Doppler broadening of the ions, which is tied to the forbidden line $z$. We allowed the Lorentzian width to vary for each line; this parameter empirically accounts for the natural line widths of the lines, as well as the representation of multiple grouped satellite lines of lower charge states with similar energies, and also a small additional contribution from the calorimeter line spread function. 
	Moreover, the line positions are allowed to vary within one FWHM to account for residual uncertainties in the transition energies, and relative intensities can vary freely to account for any differences that may stem from the polarization and cross section calculations. 
	The background is mostly due to bremsstrahlung. We used a linear function to model it in the present energy range. 
	Using these constraints, we obtained good fits to EIE spectra measured at each electron beam energy. 
	An example EIE spectrum, acquired at 2.6 keV beam energy and with medium charge balance, is shown in Fig.~\ref{fig:de26} together with our best-fit model.

	%=====================================================================
	\section{Results and Discussion}
	\label{result}
	%=====================================================================

	%=====================================================================
	\subsection{Degree of Linear Polarization}
	\label{sec:pol_result}
	%=====================================================================

\begin{figure*}
		\centering
		\includegraphics[width=\linewidth]{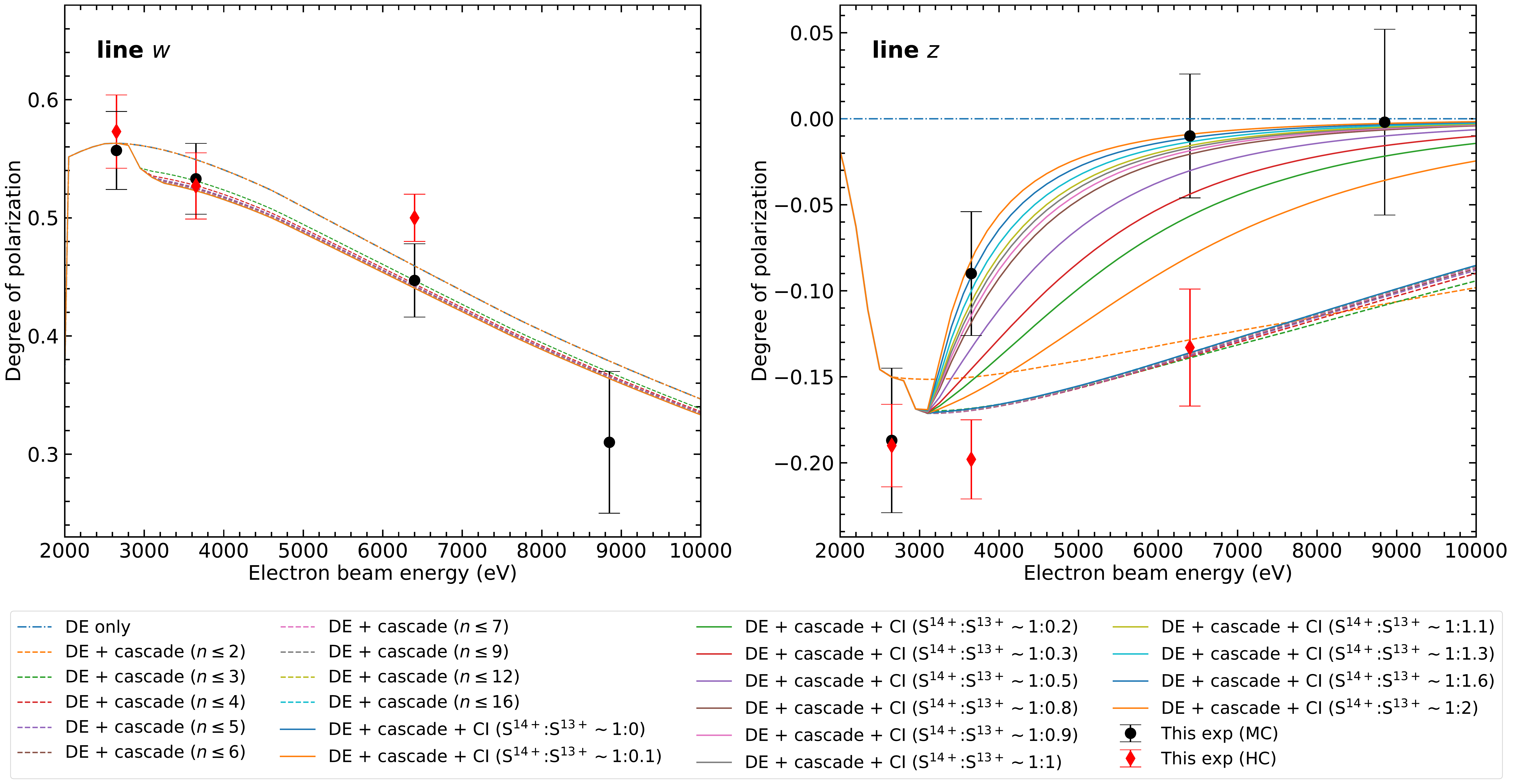}
		\caption{
			Measured degree of linear polarization for~\ion{S}{15} resonance $w$ and forbidden $z$ lines. 
			Terms MC (circles) and HC (diamonds) stand for medium charge balance and high charge balance, respectively, in our experiment. 
			Dashed-dotted line: FAC distorted-wave predictions that include only direct excitation (DE) from ground. Dashed lines: including cascades from different principle quantum numbers, from $n=2$ to $n\leq16$. Solid lines: including collisional inner-shell ionization (CI) with a varying abundance of  S$^{13+}$ with respect to that of S$^{14+}$ ions. 
		}
		\label{fig:polnmaxnp}
	\end{figure*}

	Table.~\ref{tab:results} shows the measured degree of linear polarization of lines $w$ and $z$ at four different electron beam energies.
	Our measured values for lines $w$ and $z$ agree very well with the values predicted by the distorted-wave method, within uncertainty limits (Fig.~\ref{fig:polnmaxnp}). 
	The resonance line, $w$, shows a positive degree of polarization, which suggests that it is polarized parallel to the quantization axis of the electron beam propagation direction axis.
	This is due to the fact that the $m_j = 0$ magnetic sublevel of the $1s2p \,\, ^1 P_1 $ upper state is predominantly populated relative to that of  $m_j = \pm 1$ following direct excitation from the ground state, resulting in nonzero alignment~\citep{inal1987,surzhykov2006}. 

	Conversely, for direct excitation of the $1s2s \,\, ^3 S_1 $ upper state, the populations of magnetic sublevels $m_j = 0$ and $m_j = \pm1$ are identical at any given electron-impact energy. 
	Thus, the alignment is zero, and consequently, the degree of polarization for line $z$ is zero. 
	Our basic two-level ($1s^2-1s2s$) calculation also predicts the same for line $z$ (Fig.~\ref{fig:polnmaxnp}, dashed-dotted line). 
	However, we observed a negative degree of polarization. 
	This is because the upper level of line $z$ is mostly populated by cascades from higher-lying levels~\citep{beiersdorfer1996,hakel2007,chen2015}. 
	At 2.6 keV beam energy, only cascades from $n=2$ levels are possible, and they contribute significantly to the degree of polarization of line $z$. 
	Above this beam energy, the high $1snl$ states are open for direct excitation from the ground state.
	Therefore, at 3.6, 6.4, and 8.9 keV beam energies, strong cascades from $n \geq 3$ further decrease the degree of polarization for both lines $w$ and $z$. 
	The dashed curves in Fig.~\ref{fig:polnmaxnp} show changes in X-ray polarization due to cascades from different $n$ levels, up $n=16$.

	For beam energies above 3.14 keV, collisional inner-shell ionization of lithiumlike \ion{S}{14} ions plays an important role, as it populates the $1s 2s \, ^3 S_1$ level through ionization of the lithiumlike ground state, $1s^2 2s\, ^2S_{1/2}$~\citep{inal1987}. 
	We used FAC to calculate the magnetic sublevel populations following collisional ionization, in order to check its effect on the polarization. 
	We vary the fractional abundance of \ion{S}{14} ions relative to that of \ion{S}{15} ions in these calculations. 
	The solid curves in Fig.~\ref{fig:polnmaxnp} show polarization predictions, including both cascades and CI, with a varying relative ion population. 
	The contribution to line $z$ line from CI is essentially unpolarized, as both magnetic sublevels, $m_j = \pm{1/2}$, of lithiumlike ground state $\,^2 S_{1/2}$ have identical values in an axially symmetric system~\citep{mehlhorn1968}.
	Therefore, the larger the fraction of the upper state of the line $z$ populated via CI of \ion{S}{14}, the greater the reduction in the observed degree of polarization. This effect becomes more significant with increasing electron beam energy, as the relative importance of CI increases with increasing collision energy.
	In fact, our experiment with medium charge-balance settings shows excellent agreement with the theory when CI is accounted for using the charge balance measured from the strengths of the RR peak strengths and the inner-shell satellite lines $q$ and $r$. 
	Conversely, the predicted polarization for the high charge-balance measurements has only a small correction from inner-shell ionization, and a model only including cascades agrees very well with our measurements, as the fractions of \ion{S}{14} ions are relatively small compared to those of \ion{S}{15}.

	The Breit interaction~\citep{breit1929}, a relativistic correction to the instantaneous Coulomb repulsion, is usually important for high-$Z$ ions~\citep{fritzsche2009}, but it may considerably alter the degree of polarization for low- and medium-$Z$ ions~\citep{shah2015}. 
	Moreover, such relativistic effects become progressively more significant as the incident electron-impact energy increases~\citep{reed1993}.
	We considered these effects in our calculations, but we did not find a substantial change in the degree of linear polarization for lines $w$ and $z$. 
	Moreover, two-photon $E1$ radiative decay can also affect the $1s 2l$ line polarizations~\citep{derevianko1997,surzhykov2010,dipti2020}. 
	Our level-population calculations automatically take this into account, based on two-photon rates calculated by~\citet{drake1986}.  
	The external magnetic and electric fields present in the trapping region of an EBIT could also modify the magnetic-sublevel population, thus the degree of polarization. 
	However, in our experiment, the direction of the external magnetic field is the same as that of the electron beam, so the magnetic-sublevel populations should not be redistributed.
	%
	
	%=====================================================================
	\subsection{Total EIE Cross Sections}
	\label{sec:EIE-result}
	%=====================================================================
		%

	\begin{figure*}
		\centering
		\includegraphics[width=\linewidth]{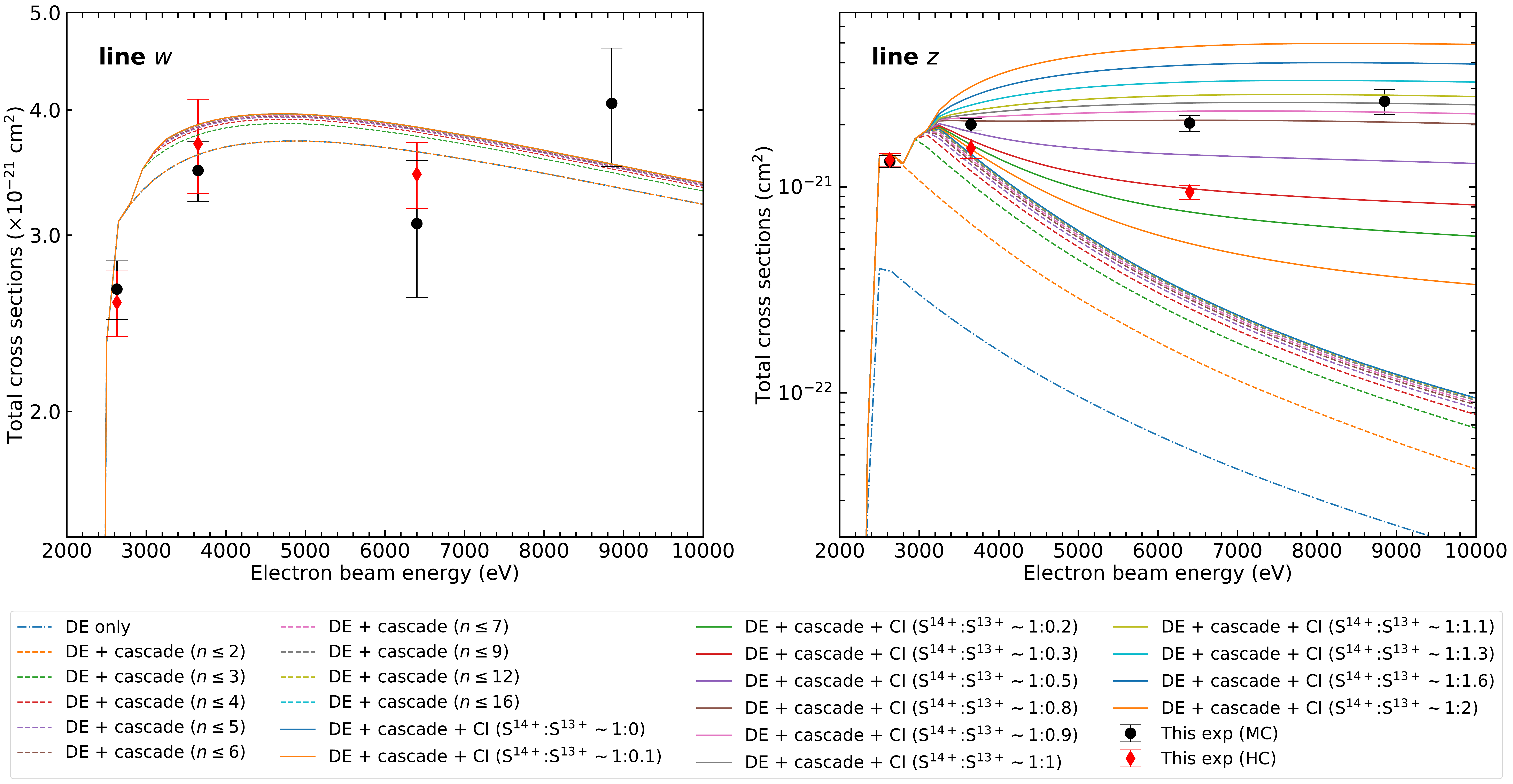}
		\caption{
			Same as Fig.~\ref{fig:polnmaxnp}, but for total line-emission cross sections.
		}
		\label{fig:crossnmaxnp}
	\end{figure*}

	The EIE cross sections at an observed angle of 90$^{\circ}$ are obtained by normalizing the EIE intensities with RR cross sections and RR intensity. 
	Finally, the 90$^\circ$ cross sections are corrected for measured polarization to extract the total line-emission cross sections. 
	The uncertainties in the derived cross sections include contributions from all possible terms, including statistical fitting uncertainties on the RR ($\sim4$\%), and EIE intensities ($\sim1$\%), detector efficiency, and filter transmissions ($\sim3$\%), theoretical differential RR cross sections ($\sim5$\%), and, as explained in Sec.~\ref{sec:pol}, the uncertainty associated with the measured degree of polarization ($\sim$5-7\%). 
	The overall uncertainty in the measured cross sections is on the order of $\sim10$\%.

	In addition, we also checked background ions with ionization energies near to that of \ion{S}{15} ions, such as \ion{O}{8} and \ion{N}{7} ions, which may contribute to the observed RR spectrum. 
	{
	In such a case, the \ion{S}{15} RR $n$=2 line can blend with RR $n$=1 lines from \ion{O}{8} and \ion{N}{7} at any given electron beam energy. }
	We cannot directly resolve these background lines in the RR spectrum, as the electron beam energy spread is much larger than the difference between the ionization potentials of background ions. 
	If present, however, the fractional abundances of such ions can be estimated based on the observed and expected strengths of collisionally-excited hydrogenlike Lyman-$\alpha$ line. 
	{In order to estimate line strengths of contaminant ions, we have removed the $12.7~\mu\mathrm{m}$ Be filter from the beam path. 
	In the case of the MC experiment, we observed only a minor enhancement over a continuum at $\sim500$ eV \ion{N}{7} Ly$\alpha$ and $\sim653$ eV \ion{O}{8} Ly$\alpha$ energies. 
	Having taken into account the ECS efficiency (Fig.~\ref{fig:QE}), we find that the data indicates a negligible contribution to the RR spectrum.}
	On the other hand, we see a strong oxygen line at 653 eV for the high charge-balance experiment. 
	This may be due to the fact that we used a pulsed injection for HC, in contrast to a continuous injection of S for the MC experiment. 
	In the latter case, the oxygen ion buildup in the trap may have been inhibited by the continuous flow of sulfur into the trap. 

	\begin{figure*}
		\centering
		%\scriptsize
		\includegraphics[width=\linewidth]{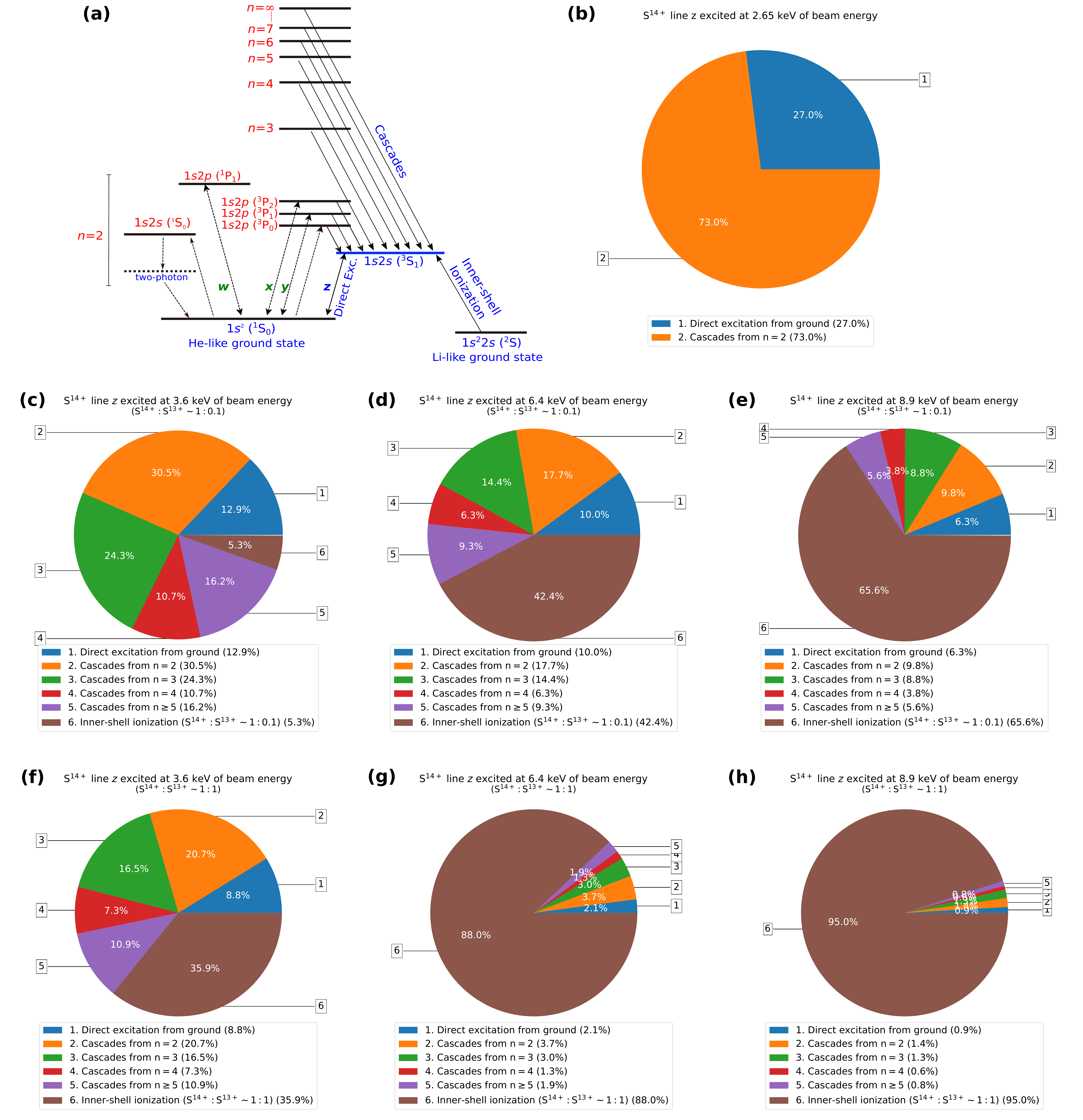}
		\caption{Top panel (a): Simplified level diagram relevant to the formation of the upper state $1s 2s \, (^3 \mathrm{S}_1)$ of forbidden line $z$ (energy levels are not to scale). Rest panels represent the relative contributions of different atomic processes to the total line-emission cross section for $z$. Top panel (b) shows contributions at 2.6 keV beam energy, below the excitation threshold of $K\beta$. Middle panels (c), (d), and (e) represent contributions at 3.6, 6.4, and 8.9 keV beam energies, where a smaller fraction ($\sim10$\%) of Li-like \ion{S}{14} is considered -- the charge balance that may exist in thermal plasma conditions. Bottom panels (f), (g), and (h) show the same as (c), (d), and (e) but for the equal fractions of \ion{S}{14} and \ion{S}{15} ions, which may exist in high-temperature out-of-equilibrium plasmas.}
		\label{fig:piecombined}
	\end{figure*}
	
	For HC data, the fractional abundance ratio between \ion{O}{8} and \ion{S}{15} can be estimated using Eq.~\ref{eq:crossEIE}, i.e., by taking the ratio between $I(\mathrm{O \, Ly\alpha})/I(\mathrm{S}{w})$~and~$\sigma_{90^\circ}(\mathrm{O \, Ly\alpha}) / \sigma_{90^\circ}(\mathrm{S}{w})$. 
	However, here we must also consider attenuation from molecular contaminants that may freeze on the filters.
	The main expected contaminants primarily contain hydrogen, oxygen, and nitrogen.
	As the photoelectric absorption cross section above threshold scales approximately as $E^3$, any such contaminant would produce the same transmission curve at energies above the oxygen $K$-edge, assuming a given optical depth at a given photon energy.
	We therefore treat the contaminant as a layer of oxygen atoms.
	We measured the optical depth decrement between two energies by taking the intensity ratio of the O Ly$\alpha$ and Ly$\beta$ lines, and comparing them to previously measured ratios using the same EBIT. 
	The O Ly$\alpha$/Ly$\beta$ intensity ratio at high impact energies becomes constant (see Fig.~23 of~\citet{beiersdorfer2003}).
	Thus, we took the weighted average of measured ratios in the range of 2--10 keV beam energies (6.13 $\pm$ 0.11). 
	Based on the optical depth decrement inferred from the measured ratio, we estimated the areal density of oxygen contaminants to be $53 \pm 6\, \mu$g cm$^{-2}$; the equivalent layer thickness was $0.53 \pm 0.06 \, \mu$m, assuming a fiducial density of 1 g cm$^{-3}$. 
	We corrected the total filter transmission for the contaminant, and obtained the relative fractional abundance between O and S ions. 
	We found these to be $\sim21$\%, $\sim38$\%, and $\sim16$\% for 2.6, 3.6, and 6.4 keV, respectively, for HC measurements.
	These oxygen fractions are used then to obtain the real RR intensity due to \ion{S}{15} ions only. 
	An extra RR line, based on the ionization potential of \ion{O}{8}, is added, and its flux is constrained to the inferred $n_{\mathrm{O}\textsc{viii}} / n_{\mathrm{S}\textsc{xv}}$ ratios in the fitting procedure, as explained in Sec.~\ref{sec:rr-exp}. 
	The inferred \ion{S}{15} RR intensity and associated uncertainty from this step are then used to obtain the total cross sections for HC measurements. 

	The final cross section results are listed in Tab.~\ref{tab:results} and shown in Fig.~\ref{fig:crossnmaxnp}. 
	Distorted wave predictions using FAC show excellent agreement with the measured cross sections, given that we account for necessary contributions from cascades and collisional inner-shell ionization, as also explained in Sec.~\ref{sec:pol_result}. 

	The measured cross sections for the resonance line, $w$, show very good agreement with the theory. 
	Cascades within $n=2$, and from high-$n$ levels, modify the $w$ cross sections by only a very small amount (6\%--7\%).
	In contrast, the cross sections for line $z$ are significantly altered by radiative cascades. 
	For example, cascades from within $n=2$ increase the cross section of $z$ by $\sim73$\% at 2.7 keV beam energy, as compared to direct excitation from the ground (Fig.~\ref{fig:piecombined}, panel (b)).  
	As the electron-impact energy increases, cascades from high $1snl$ states further enhance the emission cross sections of line $z$. 
	Figure~\ref{fig:crossnmaxnp} shows cascade contributions from different $n$ levels, up $n=16$, as dashed lines.

	In addition, as explained earlier, collisional inner-shell ionization of lithiumlike \ion{S}{14} ions starts to contribute to line $z$ above 3.14 keV beam energy. 
	As shown in Fig.~\ref{fig:piecombined}, panels (c) and (f), CI contributes as much as $\sim36$\% to the measured cross sections at a 3.6 keV beam energy, depending on the relative abundances of \ion{S}{14} and \ion{S}{15}. 
	This contribution increases significantly at high electron-impact energies, and dominates over the usual contributions from cascades and direct excitations. 
	For example, our experiment shows that CI enhances the cross sections of line $z$ by $54 ^{+0.6} _{-0.7}$\% at 6.4 keV beam energy, when the Li-like S ion concentration is increased from 24\% to 80\% in the trap. This agrees very well with distorted wave predictions of 53.4\% enhancement. 
	As depicted in Fig.~\ref{fig:piecombined}, panels (d), (e), (g), and (h), CI completely dominates the cross sections at high beam energies of 6.4 and 8.9 keV, even if the \ion{S}{14} fraction is very low compared to that of \ion{S}{15}. 
	It is therefore imperative to take CI into account for the spectral modeling of high-temperature plasmas, not only those in equilibrium conditions, but particularly those that are out of equilibrium.

	Our systematic measurements under two different charge-balance conditions clearly distinguish between predictions with different fractions of \ion{S}{14} relative to that of \ion{S}{15} and show that both cascades and CI are essential for accurate predictions of emission cross sections for line $z$. 
	Furthermore, experimental agreement between data taken at a 2.6 keV beam energy with two different charge balances indicates that the overall analysis, the extracted charge balance, as well as the theoretical treatment of cascades and CI, are reliable.

	Line $z$ may also be populated via charge exchange (CX) and RR into H-like \ion{S}{16} at higher electron beam energies, and this may modify the inferred polarization and cross sections. 
	The cross sections for RR into $1s2s$ levels are on the order of $10^{-24} - 10^{-25}$ cm$^2$ (see Tab.~\ref{tab:RR}), which are at least 3-4 orders of magnitude smaller compared to line-formation cross sections due to direct excitation, cascades, and inner-shell ionization processes. 
	On the other hand, the CX cross section are on the order of $\sigma_{\mathrm{CX}} = q \times 10^{-15}$ cm$^2$, where $q$ is the ionic charge~\citep{phaneuf1983,janev1985,otranto2006}. 
	The CX rate can be represented as the product of $\sigma_{\mathrm{CX}} \,N_{\mathrm{i}}\,N_{\mathrm{n}}\,\nu_\mathrm{in}$, where $N_{\mathrm{i}}$ and $N_{\mathrm{n}}$ are the ion and neutral densities, and $\nu_\mathrm{in}$ is the collision velocity of ions and neutrals inside an EBIT, which is set by the ion temperature.
	Previously, we measured CX between \ion{S}{16}~--\ion{}{17} and SF$_6$ using the same EBIT and ECS detector~\citep{betancourt2014}. 
	By comparing the count rate, populations of bare, hydrogenic, and heliumlike S ions, and gas injection pressure used in that experiment, as in this one, we estimated that the CX rate contributing to line $z$ is approximately three orders of magnitude smaller than the electron-impact excitation rate for our MC measurements. 
	Furthermore, our HC measurements used a pulsed gas injector to introduce neutrals to the trap only at the beginning of the charge-breeding cycle.
	Given the very low background pressure ($\leq 10^{-11}$\,Torr) at the center of the trap, and the lack of injection after the breeding part of the HC measurement cycle, X-rays from CX must be negligible.
	Data taken at 2.6 and 3.6 keV beam energies must be even less affected, due to lower production of H-like S ions, resulting in fewer targets for CX that can produce S K-shell emission lines. 
	Thus, population of the $1s2s \,\, ^3 S_1 $ level through RR and CX processes, is negligible for the present experiment. 
	Furthermore, dielectronic recombination and resonance excitation are not relevant for the present experimental electron beam energies used here, and 3-body recombination contribution should be negligible, due to the low density of the EBIT plasma~\citep{levine1988}.

	%=====================================================================
	\section{Summary and Conclusions}
	\label{concl}
	%=====================================================================
	
	We have measured the degree of linear polarization and total effective line-emission cross sections for the crucial heliumlike resonance and forbidden lines of \ion{S}{15}, using the combination of the LLNL EBIT-I facility with two high-resolution EBHiX crystal spectrometers, together with the broad-band ECS microcalorimeter. 
	Our technique enabled us to measure the emission cross sections to better than 10\%, and allowed us to disentangle the relative contributions from $n\geq3$ cascades and from inner-shell ionization of Li-like \ion{S}{14} to the key forbidden line, $z$. 
	Our systematic experiment also benchmarks FAC distorted-wave predictions for both the X-ray line polarization and total line-emission cross sections, and demonstrates that cascades and inner-shell ionization contributions should be accounted for in predictions of heliumlike $1s2l$ line-emission cross sections. 

	In astrophysical conditions where the ionization balance is not solely determined by the mean electron temperature of the plasma, collisional inner-shell ionization processes can contribute a considerable fraction of the line emission, as shown in this work. 
	Therefore, our data may also help with the identification of nonequilibrium conditions that may exist in transient X-ray sources, such as young supernova remnants, accretions shocks, and solar flares~\citep{watanabe1995,rasmussen2001,porquet2010,decaux1997,katsuda2012,suzuki2020}. 
	Such diagnostics, however, are heavily dependent on the underlying atomic data compiled in spectral modeling codes, such as SPEX~\citep{kaastra1996} and AtomDB~\citep{foster2012} spectral modeling codes. For example, the compiled collision strengths for the forbidden line in \ion{Fe}{25} in these codes critically differ by more than 40\%~\citep{hitomiatomic2018}. 
	Therefore, our experimental data could also be used to stringently test the accuracy of these codes. 
	This is indeed a crucial task in view of the next generation of X-ray satellites, namely \textit{XRISM}~\citep{xrism2018} and \textit{Athena}~\citep{barret2016}, which will reach exceptional spectral-energy resolutions and higher sensitivities, as shown by \textit{Hitomi}~\citep{hitomi2016,hitomi2017}.

%====================================================================
%======= ACKNOWLEDGMENTS =============================================
%=====================================================================
	
	\acknowledgments{
		C.S. acknowledges the support from an appointment to the NASA Postdoctoral Program at the NASA Goddard Space Flight Center, administered by the Universities Space Research Association under contract with NASA, by the Lawrence Livermore National Laboratory (LLNL) Visiting Scientist and Professional Program Agreement No.\ VA007036 \& VA007589, and by Max-Planck-Gesellschaft (MPG). {M.F.G is supported in part by NASA APRA grant No.~80NSSC20K0835.} Work by LLNL was performed under the auspices of the U. S. Department of Energy, under Contract No.\ DE-AC52-07NA27344 and supported by NASA Astrophysics Research and Analysis (APRA) Program grants to LLNL and NASA/GSFC. We also acknowledge support from NASA's Astrophysics Program. We would like to thank Klaus Widmann and Ed Magee for assisting with these measurements.
	}

%%%%%%%%%%%%%%%%%%%%%%%%%%%%%%%%%%%%%%%%%%
%\bibliographystyle{aasjournal}
%\bibliographystyle{aa_url}
%\bibliography{main_revised}

%%%%%%%%%%%%%%%%%%%%%%%%%%%%%%%%%%%%%%%%%%
\end{document}